\documentclass[useAMS,usenatbib]{mn2e}

\usepackage{graphicx}
\title[Astrochemical Evolution of Turbulent Giant Molecular Clouds]{The Astrochemical Evolution of Turbulent Giant Molecular Clouds : I - Physical Processes  and Method of Solution for Hydrodynamic, Embedded Starless Clouds}
\author[A. Kumar and R. Fisher]{A. Kumar$^{1}$\thanks{E-mail:
rfisher1@umassd.edu (RTF)} and  R. Fisher$^{1}$ \\
$^{1}$285 Old Westport Road, University of Massachusetts, North Dartmouth 02747, Massachusetts, USA}
\begin{document}

\date{Accepted January 29, 2013}

\pagerange{\pageref{firstpage}--\pageref{lastpage}} \pubyear{2011}

\maketitle

\label{firstpage}

\begin{abstract}
Contemporary galactic star formation occurs predominantly within gravitationally unstable, cold, dense molecular gas within supersonic, turbulent, magnetized giant molecular clouds (GMCs). Significantly, because the chemical evolution timescale and the turbulent eddy-turnover timescale are comparable at typical GMC conditions, molecules evolve via inherently non-equilibrium chemistry which is strongly coupled to the dynamical evolution of the cloud.  

Current numerical simulation techniques, which include at most three decades in length scale, can just begin to bridge the divide between the global dynamical time of supersonic turbulent GMCs, and the thermal and chemical evolution within the thin post-shock cooling layers of their background turbulence.  We address this GMC astrochemical scales problem using a solution methodology, which permits both complex three-dimensional turbulent dynamics as well as accurate treatment of non-equilibrium post-shock thermodynamics and chemistry.  

We present the current methodology in the context of the larger scope of physical processes important in understanding the chemical evolution of GMCs, including gas-phase chemistry, dust grains, surface chemistry, and turbulent heating. We present results of a new Lagrangian verification test for supersonic turbulence. We characterize the evolution of these species according to the dimensionless local post-shock Damk\"{o}hler number, which quantifies the ratio of the dynamical time in the post-shock cooling flow to the chemical reaction time of a given species.  

Lastly, we discuss  implications of this work to the selection of GMC molecular tracers, and the zeroing of chemical clocks of GMC cores.

\end{abstract}


\section{Introduction}

Giant molecular clouds (GMCs) are rich dynamical structures resulting from the interplay of many complex processes, including supersonic magnetized turbulence, self-gravity, chemical evolution, star formation, jets and outflows, and radiative transfer.  Because star formation occurs exclusively within gravitationally unstable cold dense gas within GMCs \citep {shuetal87, mckeeostriker07}, they play a crucial role in connecting larger-scale galactic dynamics to protostars, protostellar disks, and planet formation.  The goal of this paper is to develop novel simulation  techniques which will help elucidate the astrochemical evolution of realistic, turbulent GMCs, and in turn yield insight into both observations and theoretical models of GMCs.

\smallskip

GMCs are also host to a wide range of complex molecules, which play vital roles in both the
diagnostics and the dynamics of the cloud. Observers have detected a wide range of molecular species
in GMCs, including simple diatomic molecules like CO and CS, up to complex organic molecules like
formaldehyde (H$_{2}$CO), methanol (CH$_3$OH)  and amino acetonitrile, potentially a precursor to the
amino acid glycine. Tracer molecules like CO and NH$_3$ play a vital role as observational diagnostics
of the density, temperature, and velocity dispersion of the cloud. Additionally, the chemistry of key
ions, such as  HCO$^+$, regulate the overall ionization level of the cloud, and establish the degree to
which the magnetic field is coupled to the gas.

\smallskip

Much of the work done to understand the chemical evolution of GMCs assumes the hydrodynamic background of the cloud is either fixed or smoothly varying. Crucially, supersonic turbulence significantly modifies this quiescent picture of the chemical structure of GMCs through two key mechanisms. Firstly, shock heating generates high temperatures in post-shock cooling flow regions,  and promotes gas phase reaction pathways with high activation energies -- particularly of neutral-neutral  chemistry -- which are suppressed at lower temperatures \citep {padoanetal00}. Secondly, parcels of gas are compressed and rarefied by the turbulence, resulting in a log-normal distribution of density values \citep {padoanetal1997,vazquez1994,federrathetal08,passotvazquez1998}. This broad distribution of densities results in significant departures of local reaction rates from the mean over the entire cloud \citep {glovermaclow2007a,glovermaclow2007b}.

\smallskip

The Atacama Large Millimeter Array (ALMA) promises unprecedented observations of molecular line transitions in the millimeter-range with spatial resolution of .01 arcseconds (or roughly 1 AU at the distance of Taurus) -- ten times better resolution than either Very Large Array (VLA) or the Hubble Space Telescope (HST), and high enough to begin to peer into the detailed shock dynamics of GMCs \citep {herbst08}. The challenge to modelers is to match  such high-quality GMC astrochemical observations with next-generation theoretical and computational tools which build upon and extend existing techniques.

\smallskip

To date, the chemical evolution of GMCs has been modeled in one of two classes of approximations. The first set of approximations begin with extensive chemical  networks and incorporate simplified dynamics -- typically either spatially homogeneous or spherically-symmetric  \citep{aikawaetal05, aikawaetal08}.  The second includes a pioneering body  of work which, for the first time,  coupled  realistic three-dimensional hydrodynamical or magnetohydrodynamical turbulent simulations, with  simplified  chemical networks and molecular line cooling \citep {maclowpavlovski4,pavlovskietal06, maclowpavlovski2, maclowpavlovski3, glovermaclow2007a, glovermaclow2007b, gloveretal10}. Both approaches have clear merits, but also significant limitations. Notably, as we will discuss in more detail below, current numerical simulation techniques are limited in their ability to resolve the thin post-shock cooling layers in turbulent GMCs. Ultimately what is needed to fully understand the astrochemical evolution of realistic GMCs is a full multidimensional simulation, coupled to extensive chemical networks, and an accurate treatment of the thermodynamics of the gas in thin cooling layers. As we discuss below, this goal presents significant technical challenges, which we directly address in this paper.

\smallskip

\begin{figure*}
\centering%
\vspace{-0.5cm}
   \hspace*{2.68cm}\includegraphics[width=140mm]{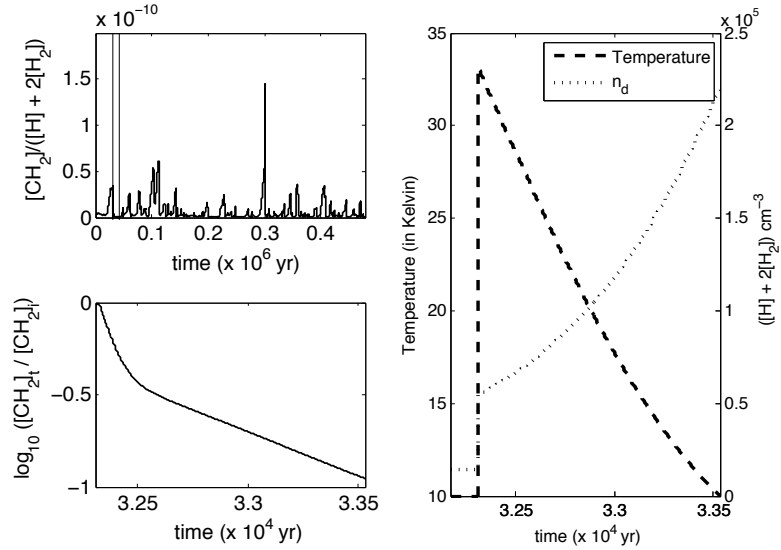}
\caption{A diagram illustrating the large dynamic range between the global dynamical time and the local thermodynamic and chemical time in post-shock cooling flows, for a representative species (CH$_2$). The figure at upper left shows the evolution of the normalized abundance of CH$_2$ over a global dynamical time, along a single fluid parcel.  The vertical lines indicate the portion of the time domain blown-up at right, depicting the temperature and gaseous number density evolution post-shock. The figure at lower left depicts the log normalized abundance evolution of CH$_2$ over this same shock.
\label{schematic}}
\end{figure*}

Modeling the post-shock cooling region while also fully simulating the global dynamics of a turbulent supersonic clump presents a significant challenge to any 3D simulation, even with the power of adaptive mesh refinement. The postshock cooling region behind a J-shock is of order $\sim$ 1 AU in thickness at $n = 10^5$ cm$^{-3}$ for a shock velocity $v_s \sim 10$ km/s, and behind a C-shock of order $\sim$ 10 AU thickness under similar conditions \citep {shullhollenbach78}. In contrast, a supersonic GMC clump of $10^3 M_{\odot}$ is  of order 1 pc in size, resulting in a dynamic range of four to five decades in length scale between the largest and smallest scales required to simultaneously model both supersonic turbulence and molecular shock chemistry. This dynamic range is roughly one to two decades greater than even the most highly-resolved simulations of turbulent GMCs \citep {kritsuketal06, kritsuketal07, federrathetal10}.  This large dynamic range of timescales is depicted in figure \ref {schematic}, which shows both the global chemical evolution of a representative species in a turbulent GMC clump along a single parcel, as well as its local hydrodynamic and chemical evolution through the post-shock flow of a single J-shock. 

 Significantly, for a typical supersonic GMC clump simulation with 3-D RMS Mach number 3.5   on an Eulerian 512$^3$ mesh, the timestep is roughly 500 yr, assuming a Courant number of 0.5,   implying the entire post-shock flow is only captured within $\sim$ 2 timesteps. Consequently, the thickness of shocks is unresolved in multidimensional large-scale GMC simulations. This is a significant limitation to understanding the global astrochemical evolution of a GMC, since non-equilibrium cooling and chemistry become significant in the post-shock cooling region where molecular line cooling determines the post-shock temperature. 

\smallskip

We address the GMC astrochemical scales problem using a novel solution methodology, which permits both complex three-dimensional turbulent dynamics as well as an accurate treatment of non-equilibrium post-shock chemistry.  The key idea is to separate the dynamical and chemical evolution into two weakly-coupled problems, and treat the wide range of scales between the dynamical time on the scale of a GMC clump $t_{\rm dyn} \sim L / v \sim 10^6$ yr  to the cooling timescale in a thin post-shock layer, $t_{\rm cool} \sim 10^3$ yr. This decoupling of the dynamics from the molecular cooling and chemical evolution in the post-shock flow is a valid starting point for an approximation because dust grain and molecular line cooling are highly efficient in establishing the temperature of the cloud,  and the shock jump conditions themselves are simply fixed by conservation.  Indeed, detailed three-dimensional simulations of turbulent GMCs including chemical, kinematic, and thermodynamic evolution in a GMC have demonstrated that the equation of state is well-approximated as isothermal everywhere outside of the immediate radiative post-shock cooling layers  \citep {pavlovskietal06}.  Consequently, the large-scale supersonic hydrodynamical evolution can be accurately treated using an isothermal approximation to the Euler equations of hydrodynamics, while the detailed molecular and atomic cooling and chemical evolution in thin post-shock flows and in regions of intense shear dissipation   can be treated separately in post-processing.

\smallskip

This paper is organized as follows. In \S \ref {physicalprocesses}, we present a comprehensive overview of the primary physical processes which govern the physical and chemical evolution of a GMC, including gas-phase chemistry, dust grains, surface chemistry, and turbulent heating. In \S \ref {methodologysection}, we describe in detail the numerical methodology developed to treat the physical and chemical evolution within supersonic isothermal turbulent GMCs, including gas dynamics, driven turbulence, Lagrangian tracer particles, gas-phase chemistry, gas-phase cooling, and J- shock modeling. Section \S \ref {scaling} presents the dimensionless parameterization selected for our hydrodynamics models. Section \S \ref {verification}, we present new verification tests developed to test the new aspects of the numerical methods  here.  In section \S \ref {resultssection} , we present preliminary results for the astrochemical evolution along a sample Lagrangian trajectory as well as a ensemble of Lagrangian trajectories calculated with our new numerical methods. We calculate the joint PDF of density and temperature, and infer the mass filling fraction of the warm molecular phase generated by turbulent heating within our models. Lastly, in section \S \ref {discussion}, we discuss our findings and conclude.

\section {Physical Processes}
\label {physicalprocesses}

\subsection {Gas-Phase Chemistry}

In a key paper, \cite{herbst1973} demonstrated that ion-neutral chemistry could provide
formation channels for a number of molecular species, such as OH and H$_2$O. Specifically, cosmic rays
are able to ionize molecular hydrogen even deep in the interior of GMCs, and thereby initiate ion-neutral
chemical reactions. The classical Langevin rate of these ion-neutral reactions is independent
of velocity, and therefore able to proceed even at the low mean background temperatures within
GMCs. Consequently, the ion-neutral gas-phase molecular chemistry timescale is set by cosmic ray
injection in dense GMCs and is independent of density \citep{herbst1973,bergintafalla07}, 
with a characteristic chemical time scale $t_{\rm chem}$ yr.  

\smallskip

In addition to ion-neutral chemistry, neutral-neutral chemistry is also now understood to play an
important role in the chemical evolution of GMCs even at low temperatures. Direct measurements
of some neutral-neutral reaction rate have demonstrated they they have relatively rapid reaction
rates ( $>$ 10$^{-10}$ cm$^3$ molecule$^{-1}$ s$^{-1}$) even at low temperatures down to 10 K \citep{smith1997}. In addition, other neutral-neutral reactions can be activated in warm regions with sufficient gas kinetic temperature to overcome activation barriers. Our initial models  include only a few dozen simple carbon, nitrogen, and oxygen- bearing species (including, among others, C, CH$^+$, CO, HCO$^+$, O, O$_2$, OH, H$_{2}$O, N, and NH$_3$ - see table \ref{tab:speclist} for a complete list) for which the rate reactions are well-determined.

\smallskip

A realistic treatment of both ion-neutral and neutral-neutral gas phase chemistry is at the heart of
this paper. To further the state-of-the-art in astrochemical modeling, we have developed a new astrochemistry framework, ASTROCHEM, which builds upon and extends existing astrochemical codes
and databases using a sophisticated automatic chemical network pre-processor, as discussed in more detail in our methodology section (\S \ref {gasphasechemistry}). ASTROCHEM allows for the efficient and flexible solution of large chemical networks, as discussed below  in our methodology section. The chemical reaction network presently included in ASTROCHEM have been tabulated in appendix \ref {chemeqtable}, along with the mathematical expression of reaction rate
coefficients of reactions and their references.

\subsection {Dust Grains and Surface Chemistry}

In addition to the gas phase molecular species, dust grains also play a crucial role in the chemistry
and thermodynamics of GMCs. Because hydrogen has a high surface mobility on dust grains,
a key feature of dust grain chemistry is the hydrogenation of molecules, providing an important
formation mechanism for complex organic molecules including methanol \citep{dishoeckblake1997}.
Although there is a significant body of both theoretical and experimental work on the
surface chemistry at conditions appropriate for GMCs (e.g., recent review by \citealp {herbstetal2005}
among others), the complexity of the subject is significant. In particular, the reaction rates depend
directly on the surface mobility of hydrogen and other light molecules, which in turn depends on
complex interfacial physics, including whether the surface molecules are strongly bound to the
surface through chemisorption or weakly bound through physisorbtion. There are also outstanding
research issues with regard to the methodology used to evolve surface chemical species; while rate
equation methods are still widely used, this approach formally breaks down for trace species for
which the mean abundance may be less than a single molecule per grain.

\smallskip

The chemistry models present in this paper consist solely of gas-phase chemistry, and include
the effect of the dust grains on the molecular hydrogen chemistry and gas heating/cooling only.
This model will later be extended to incorporate a grain surface chemistry reaction network (e.g.,
Herbst et al, 2005) using a rate reaction formalism built upon ASTROCHEM.

\subsection {Turbulent Heating}

Turbulence has long been recognized as playing an important role in the dynamics of GMCs, at
least to {Larson}'s seminal paper on the internal velocity dispersion in molecular clouds \citep{larson81}.
Turbulence decays on the order of a dynamical time \citep{maclow1999}, thereby leading to a significant source of heating for GMCs. This idea was initially explored in the context of atomic HI clouds. Earlier work has discussed the impact of subsonic turbulence on the gas kinetic temperature of atomic HI clouds \citep {falgaronepuget1995}. Extensions of this work using a simplified model of intermittent turbulent heating demonstrated that localized hot regions could form within otherwise cold HI clouds, thereby activating neutral-neutral gas phase chemical reactions which were suppressed at the background temperature  \citep{falgaroneetal1995}.

\smallskip

More recently, simulations of molecular gas in turbulent GMCs have demonstrated that turbulent heating alone can largely account for the mean temperature within GMCs \citep{pan2009}. Because of the intermittent nature of the turbulent heating, post-shock temperatures in excess of 1000 K can be reached in simulation models \citep{pavlovskietal06}. Strong shears may also lead to a large dissipation rate; the specific turbulent heating rate per unit mass is relatively uncorrelated with density \citep{pan2009}. However, the majority of the mass of the turbulent GMC model is concentrated within shocks. Extensions of the She-Leveque model of turbulent intermittency reveal the dimension of the most dissipative structures in supersonic turbulence which best fit both numerical and observational data to be  corresponding to shocks, and not  corresponding to vortices, as is the case with subsonic turbulence \citep{boldyrev2002,pan2009}.  Recent work has  shed more light on intermittency models and fractal dimensions of the most dissipative structures of supersonic turbulence \citep {schmidtetal2008,konstandinetal2011}. 

\smallskip

A major goal of the current paper is to devise a consistent numerical methodology which allows us to treat both turbulent shock heating and shear dissipation in supersonic isothermal GMCs with high temporal and spatial resolution. As we will describe,  our numerical simulations capture the shocks on the mesh, allowing us to include the effects of shock heating  on subgrid scales using post-processing of Lagrangian tracer particles. Furthermore, we also take into account dissipation of shear energy on resolved scales. We discuss the methodological procedures which allow us to incorporate these effects more fully in section \S \ref {methodologysection}.

\subsection{Characteristic Length Scales and Dimensionless Numbers}
\label {dimensionlesssubsection}

We begin by identifying the key length and time scales relevant to the supersonic chemical evolution of the GMC. Key physical insight into the physical and chemical evolution of a GMC can be obtained by examining the characteristic time scales of the GMC, as well as the dimensionless numbers characterizing its properties.  Whereas recent work has elucidated the dimensionless numbers important in passive scalar mixing in isothermal supersonic turbulence  \citep{scan2010}, we focus on the characterization of the chemical evolution of the GMC. 

\smallskip

The Reynolds number $\rm{Re}$ quantifies the ratio of inertial to viscous forces, and is defined on the length scale $l$ as:
\begin{eqnarray}
{\rm Re} \sim \frac{v(l)l}{\nu} \sim 7.49 \times 10^8 \left ( \frac{l}{1 \ {\rm pc}} \right ) \nonumber \\
\times \left ( \frac{v}{ 1.22  \ {\rm km} {\rm s}^{-1}} \right ) \left ( \frac{\nu}{5.00 \times 10^{14} \ {\rm cm}^2 \ {\rm s}^{-1}} \right )^{-1},
\end{eqnarray}
where $v$ is the turbulent RMS velocity, and  $\nu$ is the kinematic viscosity due to neutral-neutral molecular collisions (defined in equation \ref {kinematicviscosityscaling} in appendix
 \ref {microscopicphysics}).  The Reynolds number within GMCs is enormously larger than critical value for the transition to fully-developed turbulence measured in terrestrial experiments 
\citep{pope2000}, and so we expect GMCs to be fully turbulent. A wide variety of observational diagnostics, including power spectra measured in both the velocity and density fields, 
are consistent with this expectation \citep{scaloelm2004a, scaloelm2004b,maclowklessen2004}. 

\smallskip

The Kolmogorov scale $\eta_{\rm K}$ is the length scale where the turbulent kinetic energy is ultimately dissipated due to viscous forces. Mathematically,
\begin{eqnarray}
\eta_{\rm K} \sim \left ( \frac{\nu^3}{\epsilon} \right )^\frac{1}{4} \sim 3.49 \times 10^{-2} {\rm AU} \left ( \frac{\nu}{5.00 \times 10^{14} \ {\rm cm}^2 \ s^{-1}} \right ) ^{\frac{3}{4}}\nonumber \\
  \left ( \frac{\epsilon}{2.02 \times 10^{-4} \ {\rm erg} \ {\rm cm}^{-3} {\rm g}^{-1} {\rm s}^{-1}} \right )^{-\frac{1}{4}} ,
\end{eqnarray}
\noindent where $\epsilon$ is the specific turbulent kinetic energy dissipation rate, and $\nu$ is the kinematic viscosity. For simplicity, we have assumed that $\epsilon$ is independent of the length scale;
here it has been calculated at the integral scale using fiducial values for a representative Mach 3.5 clump. Equivalently, the Kolmogorov scale is the length scale where the Reynolds $\rm{Re}$ number is one. The eddy turnover time at the Kolmogorov
scale is the local Kolmogorov time scale $t_{\rm K}$,
\begin{eqnarray}
t_{\rm K} \sim  \left ( \frac{\nu}{\epsilon} \right )^\frac{1}{2} \sim 29.2 \  {\rm years} \times \left ( \frac{\nu}{5.00 \times 10^{14} \ {\rm cm}^2 \ s^{-1}} \right ) ^{\frac{1}{2}} \nonumber \\
\times  \left ( \frac{\epsilon}{5.89 \times 10^{-4} \ {\rm erg} \ {\rm cm}^{-3} {\rm g}^{-1} {\rm s}^{-1}} \right )^{-\frac{1}{2}} .
\end{eqnarray}

\noindent Both $\eta_{\rm K}$ and $t_{\rm K}$ are vastly smaller than the computational grid size $\Delta x$ and timestep $\Delta t$, respectively, in 3-D simulations. Consequently, any direct numerical simulation approach to turbulent mixing is computationally prohibitive, and some effective modelling is required.

\smallskip

The Knudsen (Kn) is the ratio between molecular mean free path and a representative physical length. This determines the validity of the continuum hypothesis. If the Knudsen number is $<< 1$ , then the mean-free path length is significantly less than the typical flow scale and the continuum approximation is valid. The mean-free path length of the ion and neutral species are, however, different, so their Knudsen numbers vary. The neutral-neutral Knudsen number is given by:
\begin{eqnarray}
{\rm Kn_{nn}} = 1.37 \times 10^{-9} \left ( \frac{n_{\rm n}}{4 \times 10^4 \ {\rm cm}^{-3}} \right )^{-1}  \nonumber \\
{\left ( \frac{\sigma_{\rm in}}{6.07 \times 10^{-15} \ {\rm cm}^2} \right )^{-1}\left ( \frac{L}{1 \ {\rm pc}} \right )^{-1}}.
\end{eqnarray}

\noindent The Kolmogorov length scale sets the smallest length relevant to turbulent flows. For a Kolmogorov length scale of $\sim 4.6 \times 10^{-2} $ AU, the neutral-neutral Knudsen number $\rm{Kn_{nn}}$ is $\sim 6.1 \times 10^{-3}$. Consequently the fluid approximation remains valid throughout the turbulent flow, even far beneath the length scales currently probed by numerical simulations.

\smallskip

Next, we compare the turbulent eddy turnover timescale with the timescale required for ion-neutral
molecular chemistry. In reactive chemical flows, the Damk\"{o}hler number {\rm Da} is defined as the ratio of $t_{\rm eddy} (l)$ to $t_{\rm chem}$. For {\rm Da} $ \gg 1$, chemical equilibrium is reached on an essentially frozen background flow. In contrast, for {\rm Da} $\leq 1$ (the ``well-mixed reactor'' regime), a turbulent eddy can turn over before ion-neutral chemical equilibrium is established. Consequently, for {\rm Da} $\leq 1$, turbulence plays an essential role in determining the final chemical products. We define the turbulent eddy turnover time timescale $t_{\rm eddy}$ as:
\begin {equation}
t_{\rm eddy} (l) \simeq {l \over v(l)} \simeq \left( {l L \over v_0^2}\right)^{1/2},
\end {equation}
\noindent where we have made use of Larson's law, $v(l) = v_0 (l / L)^{1/2}$, which specifies the scaling of the turbulent RMS velocity $v_r$ with the length scale $l$ \citep {larson81}.
Here $v_0$ is the RMS velocity on a characteristic scale $L$; we adopt $v_0 = $\ 1 km/s at $L = $ 1 pc as fiducial values for galactic GMCs.

The ion-neutral gas-phase molecular chemistry timescale is set by cosmic ray injection in dense GMCs
and is independent of density \citep {bergintafalla07} -- $t_{\rm chem} \sim 3 \times 10^5\ {\rm yr}$.

\smallskip
For these fiducial values of galactic GMCs, the Damk\"{o}hler number established by turbulence transitions through Da = 1 at a characteristic scale of roughly 1 pc. Using Larson's mean density-size relation, and scaling to a fiducial mean density of $10^5$ cm$^{-3}$ at 0.1 pc, we find that this characteristic length scale corresponds to a mean density of $10^4$ cm$^{-3}$, which is characteristic of the density
of large-scale clumps in galactic GMCs. On smaller length scales, the chemical evolution timescale and the eddy turnover timescale are comparable, and molecules evolve via inherently non-equilibrium chemistry which is strongly coupled to the dynamical evolution of the cloud. Therefore on the clump scale or smaller, the timescale for turbulent mixing is comparable to the ion-neutral chemical timescale. Consequently, on the clump scale or smaller, dynamical processes play a essential role in determining GMC chemical abundances.

\smallskip

\section {Methodology}
\label {methodologysection}

\smallskip

In the present work, we model the turbulence within the GMC as purely hydrodynamic, or equivalently, that the clouds are magnetically supercritical. Consequently, all shocks are modeled as J-shocks.  While there is indirect theoretical evidence that GMCs are magnetically supercritical with respect to their mean magnetic field \citep {mckeeostriker07}, there is an absence of direct measurements constraining the mass-to-flux ratio on large scales.  Future work will build upon the current framework to incorporate the magnetic field and ambipolar diffusion, and treat C-shocks.

\smallskip

In addition to the effects of shock heating, we also take into account the solenoidal heating due to dissipation within vortices and strong shears.  In contrast to previous work which includes subgrid modeling of this effect \citep{falgaronepuget1995, pan2009}, we account for the solenoidal heating on resolved scales only. Even on these resolved scales, the effect of solenoidal heating is significant, and accounts for two-thirds of the total turbulent heating in our models (\S \ref {solenoidalheating}).

\smallskip

An extensive body of simulations has demonstrated that undriven, supersonic turbulence decays rapidly on a dynamical timescale \citep{maclow1998}.  Furthermore, the statistical properties of observed giant molecular clouds cores have been demonstrated to be in better agreement with driven, steady-state turbulence than with undriven, decaying turbulence \citep{offner2008}. Consequently, we model turbulent giant molecular clouds as being driven over a narrow band of wavenumbers. On smaller scales, an inertial range is established. Recent work show that forcing of the turbulence significantly affects the density and velocity statistics of supersonic turbulence in both the inertial and injection range \citep{schmidtetal2008,federrathetal10,konstandinetal2011}. Since we post-process Lagrangian tracers by detecting shocks to get chemical evolution in our paper, any change in density scaling will change the way our chemical network evolves.

\smallskip

Our solution methodology consists of several fundamental ingredients. The first involves Eulerian hydrodynamics (\S \ref {gasdynamics}). The Eulerian hydrodynamical flow is driven on large scales to establish a turbulent power spectrum (\S \ref {driventurbulence}). The Eulerian calculation also has embedded within it Lgrangian tracer particles (\S \ref {lagrangiantracers}) which follow the flow and capture the Lagrangian state of the system for later detailed gas-phase chemistry (\S \ref {gasphasechemistry}), and radiative shock chemistry post-processing (\S \ref {postprocessing}).

\subsection {Gas Dynamics}
\label {gasdynamics}

We solve the Euler equations of hydrodynamics with driving source source terms,

\begin{equation}
 {\partial \rho \over \partial t} + \vec {\nabla} \cdot (\rho \vec {v} ) =
0,
\end{equation}

\begin{equation}
{\partial (\rho \vec {v}) \over \partial t} + \vec {\nabla} \cdot (\rho \vec
{v} \vec {v} ) = - \vec {\nabla} P +  \vec{f},
\end{equation}

\begin{equation}
{\partial (\rho E) \over \partial t} + \vec {\nabla} \cdot [ (\rho E + P)
\vec {v}] =  \vec{v} \cdot \vec{f},
\end{equation}

\begin{equation}
P = (\gamma - 1) \rho e.
\end {equation}

Here $\rho$ is the mass density, $\vec {v}$ is the velocity, $P$ is the gas pressure, $E$ is the total energy density (related to the internal energy per unit mass, $e$, by $E = e + {1 \over 2} v^2$). $\vec {f}$ is a force density representing the forcing of the fluid on large scales to sustain driven turbulence against decay (see \S \ref {driventurbulence}).

\smallskip

We model isothermal driven turbulence with an adiabatic equation of state with a ratio of specific heats $\gamma \simeq 1$. An isothermal gas can be thought of as an adiabatic gas in the limit that the number of internal degrees of freedom approaches infinity, and can be approximately modeled by taking a large but finite number of degrees of freedom. This approximation is easily accommodated by many existing hydrodynamics solvers, and has a domain of validity provided that the total heat energy added to the system over the duration of the simulation is much less than the initial internal energy.  However, the specific turbulent heating rate, which scales as $ v^{2} / (v/L) \sim{v^{3}/L} \sim{ {\cal{M}}^{3} }$,  places a strict upper-bound constraint on the choice of $\gamma$ for supersonic turbulence.

\smallskip

Consider a supersonic turbulence simulation modelled using an adiabatic equation of state in a periodic domain. The initial specific internal energy is fixed by the isothermal sound speed and the ratio of specific heats, and is simply $e_{\rm int}$ = $ c_{\rm iso}^2 /(\gamma - 1)$. Over each dynamical time, the decay of turbulent energy will deposit  $\sim 1/2 {\cal M}^2 c_{\rm iso}^2$ of heat energy per unit mass into the system. Physically, this energy is radiated from an isothermal system, but under the adiabatic approximation, the energy is simply deposited as heat. Consequently, in order for the gas to remain approximately isothermal over a total evolution of $N_{\rm dyn}$ dynamical times, we require

\begin {equation}
\gamma - 1  \ll  {2 \over N_{\rm dyn} {\cal M}^2 }  
\end {equation}

In the driven simulations presented here, one begins with quiescent gas and must evolve for a minimum of $N_{\rm dyn} \sim 2 - 3$ to reach a steady-state. Typically, in order to gather sufficient turbulent statistics in steady-state, one wishes to run for $N_{\rm dyn} \sim 10$. We find that for modelling a typical turbulent GMC clump-sized region, with ${\cal M} = 5$,  and  $N_{\rm dyn} \sim 10$,  we require $\gamma - 1 \ll .008$.  In the simulations presented here, we utilize $ \gamma = 1 + 10^{-6} $, which produces excellent steady-state results. While higher values of $\gamma$ are found in literature (often as large as $ \gamma = 1 + 10^{-2}$), these will not be sufficient to maintain isothermality throughout the course of a highly-supersonic simulation, and will lead to non-steady turbulence statistics.

 \subsubsection{Self-Consistent Choice of Mean Temperature and Mach Number}

The Eulerian frame calculations are completed using isothermal
hydrodynamics, and can be characterized by a single dimensionless
number, the 3-D RMS turbulent Mach number. The thermal evolution of
Lagrangian fluid elements must further incorporate an energy equation
including both heating and cooling, and requires an initial
temperature. After passing through shocks, Lagrangian fluid elements
return to the initial temperature. An additional self-consistency
requirement is therefore that the mean temperature within the Eulerian
volume is equal to the mean temperature along Lagrangian particles.
This is equivalent to asserting that the overall system is in net
thermal equilibrium, once all cooling and heating processes, including
turbulent heating, are taken into account.

\smallskip

In order to set the mean temperature self-consistently for a given
Mach number, we start with an initial guess for the equilibrium
temperature. We then iterate by finding the equilibrium temperature
achieved by matching the net gaseous phase molecular cooling  \citep {goldsmith2001} with net heating, including turbulent heating function
 \citep {pan2009}, and dust grain-gas cooling \citep {hollenbach1989}. During the
calculation, we maintain the dust grains to be at 10 Kelvin, as dust
grains can radiate away heat very efficiently at the GMC densities. An
updated Mach number is computed at an isothermal sound speed
reflective of the new mean temperature. We then iterate until we reach
convergence, ending with a self-consistent choice for both the
temperature and Mach number, which will in general both differ from
their initially assumed values. The converged values for both the mean
temperature and the Mach number are then used for the Eulerian
calculation. For the simulations presented here, we have employed a self-consistent mean  temperature of 14.8 Kelvin at a 3D RMS  Mach number of 3.5.

\subsubsection {Scaling}
\label {scaling}

\smallskip
Our simulated models are cast in dimensionless units such that the problem domain size $ L = 1 $ in each of the three spatial dimensions, the isothermal sound speed $ \rm{c_{iso}} =1$, and the total mass $ M = 1$.  In this formulation, only one dimensionless parameter - the 3D RMS Mach number $ \cal{M}$ - fully specifies the hydrodynamics of the problem for a given turbulent realization. 

\smallskip
It is convenient for reference to scale the dimensionless problem to physical units using fiducial values. In order to facilitate this conversion, we utilize the constraint that GMCs are observed to be in approximate virial equilibrium between turbulent kinetic energy and gravitational binding energy. We define the virial parameter $\alpha_{\rm virial}$  of a spherical cloud of mass $M$ and radius $R$ to be the ratio of its turbulent kinetic energy to gravitational binding energy : $ \alpha_{\rm virial} =  5  \sigma^{2} R^2  / (2 G  M)$  \citep {bertoldi1992}, where $ \sigma^{2} =  {\cal{M}}^{2}  c_{\rm iso}^{2} / 3$ is the 1-D velocity dispersion. We generalize the definition to a uniform periodic domain by setting $L = 2 R$,

\begin{equation}
\alpha_{\rm virial} = \frac{5   {\cal{M}}^{2} {c_{\rm iso}}^{2} L}{6\  G  M}
\end{equation}

\smallskip
Once we require that the cloud be in approximate virial balance with $\alpha_{\rm virial} = 1$, this dimensionless model can be  rescaled  to values  of direct astrophysical relevance. We can perhaps most easily see this by fixing fiducial values for the isothermal sound speed $c_{\rm iso}$ and the mean cloud density $\rho_0$. The domain size $L$ then becomes

\begin{equation}
L = \sqrt{\frac{5  {\cal{M}}^{2}  c_{\rm iso}^{2}}{6  G  \rho_0  \alpha_{\rm virial}}},
\end{equation}
where $\rho_0$ is the mean cloud density, and the total mass $M = \rho_0 L^3$. The 3-D RMS velocity $v_0$ on the domain scale is then simply ${\cal M}\ c_{\rm iso}$. 
  
\smallskip
In this paper, we choose to scale to a mean density typical of GMC clumps, with number density $4 \times 10^{4} $ cm$^{-3}$ and mass density $\rho_0 = 1.28 \times 10^{-19} \ {\rm gm} \ {\rm cm}^{-3}$. This density corresponds to mean column density $\Sigma_0 = \rho_0 L =$ 0.16 gm cm$^{-2}$. We summarize the fiducial scalings used in table \ref{scalings}.

\smallskip
\begin{table}
\caption{Fiducial Scalings Used in this Paper}
\label {scalings}
\begin{tabular}{|c|c|}
\hline
3-D RMS Mach number $\cal{M}$ & $ 5 $ \\
\hline
Isothermal sound speed $c_{\rm iso}$ & $ 2.3 \times 10^4 \quad {\rm{cm}}/{\rm{s}}  $   \\
\hline
3-D RMS velocity $v_0$ & $ 8.07 \times 10^4 \quad {\rm{cm}}/{\rm{s}} $ \\
\hline
Domain size $L$ & $ 0.405\ {\rm{pc}}  $ \\
\hline
Domain mass $M$ & $ 10^2 $    $ M_{\bigodot } $  \\
\hline
Dynamical time $L / v_0$ & $  0.43 \ {\rm Myr}$ \\
\hline
\end{tabular}

\end{table}

\subsection {Driven Turbulence}
\label {driventurbulence}

\smallskip
Because turbulence is an inherently dissipative phenomenon, we drive the simulation with momentum and energy forcing terms in order to achieve a statistically-steady state. The stochastic driving method used in our simulation is the same as originally proposed by  \cite {eswaranpope88}. The turbulent velocity fluctuations are described by Fourier transform from the spatial domain. For each stirred mode of the velocity field, an acceleration is applied at each time step. The field consists of three complex phases,
with each acceleration mode evolved by a Ornstein-Uhlenbeck (OU) random process which is analogous
to Brownian motion in a viscous medium.  An OU process is defined to be a time-correlated time sequence, and has the properties of having a zero mean, and a constant root-mean-square in time. Each step in the sequence begins with the previous value, adds a Gaussian random variable  with a given variance, weighted by a driving factor $ \sqrt{1 - f^2} $  where  $ f = \exp \left (- \Delta t / \tau_{\rm cor}\right) $ then decays the previous value by a  factor $ f $.  Here, the OU process represents a velocity forcing, and  its variance is chosen to be the square root of the specific energy input divided by the decay time $ \tau_{\rm cor} $. In the time limit that the time step  $\Delta t \rightarrow 0 $, the algorithm represents a forcing term which  is a linear weighted summation of the old state with the new Gaussian random variable.

\smallskip
%At each time step, the solenoidal components of the acceleration field are projected out in Fourier space. However, because the density field in supersonic turbulence is highly inhomogeneous, the net forcing is non-solenoidal. 

At each timestep, the velocities are then converted to their physical space by a direct Fourier transform, adding the trigonometric terms explicitly. Since the stirring is done in the low-order modes, most driving involves a fairly small number of modes. Therefore this decomposition is more efficient than a fast Fourier transform. The range of modes over which turbulence is driven includes wavelengths 1/4 to 1 times the box size in the simulations presented here.   In the turbulence literature, in which the box length is set to $2 \pi$, this range of scales corresponds to wavenumbers $k$ = 1 to 4. We note that, contrary to previous claims, the acceleration field in the FLASH default turbulence driving module is non-solenoidal,  with about a 10\% admixture of power in compressional modes.

\subsection {Lagrangian Tracer Particles}
\label {lagrangiantracers}
A simulation built upon the Eulerian framework of hydrodynamics discretizes the volume of the fluid into individual spatial cells with some finite width $\Delta x$. The Eulerian frame necessarily artificially mixes fluid parcels with different histories with a resolution-dependent mixing process when they enter into the same cell. In contrast, a Lagrangian computational framework differs from the Eulerian framework in a basic yet essential regard :  individual fluid elements are followed as a function of time $t$, without any artificial mixing. The Lagrangian framework is therefore ideally suited for the purpose of following the chemical evolution of the fluid without the artificial mixing imposed by the Eulerian mesh. Any mixing in the Lagrangian framework must be introduced explicitly, which greatly facilitates a controlled treatment of this complex process.

 Passive Lagrangian tracers are characterized by their kinematic data alone : their positions $\vec {x}_i (t)$, and their velocities $\vec {v}_i (t)$, where $i$ is an index which runs over the particle number, fully specify their evolution. In addition to this kinematic data, the particles also retain hydrodynamic data, including the density $\rho (t)$ captured from interpolation from the Eulerian mesh, for use in chemical post-processing. Significantly,  Lagrangian tracers can be thought of as massless, passive particles. Consequently, while tracers are advected by the flow, they do not directly influence the velocity or the pressure of the fluid, and therefore do not back-react upon the flow.

The particle positions $\vec {x}_i (t)$ are governed by the time integration of their velocities $\vec {v}_i (t)$ obtained by interpolation from the mesh. The numerical implementation of the tracers therefore hinges directly upon both the interpolation scheme and the time-integration scheme, which we now detail.

\smallskip
\subsubsection{Particle Interpolation from Eulerian Mesh}
The software suite FLASH provides various interpolation methods for Lagrangian tracer particles to obtain hydrodynamic data from the Eulerian mesh \cite{dubeyetal2009,fryxelletal2000}. Volume-averaged Eulerian physical quantities are discretized onto the mesh in a cell-centered fashion. For example, a continuous scalar field $f (x)$ in one dimension is discretized in a given cell $i$ as:  

\begin{equation}
f_i(x) = \frac{1}{\Delta x}  \int_{x_{i-1/2}}^{x_{i+1/2}} f(x \prime)dx\prime .
\end{equation}

where $i$ is the cell index, and $\Delta x $ is the cell size. The cell extends over the interval from $x_{\i - 1/2}$ to $x_{i + 1/2}$. An approximation to the function $f(x)$ is obtained after reconstructing the profile of the scalar field within a mesh cell. For the Piecewise Parabolic Method (PPM) method, the reconstructed function is second-order accurate spatially \citep{colellawoodward84}.  In this work, we employ a quadratic interpolant scheme to achieve the same order of accuracy for Lagrangian tracer particles.

\smallskip
\subsubsection{Particle Time-Integration Scheme }
The time integration for passive particles in our simulation is implemented by Heun's method, a two-stage Runge-Kutta explicit integration scheme. For simplicity, again we present the algorithm in one spatial dimension. We represent the fluid velocity at position $x$ and time $t$ by $v(x,t)$. Because the Eulerian mesh retains only discretized information of the fluid velocity, $v (x, t)$ is obtained by interpolation from the mesh to the location of the particles as just discussed. We therefore implement Heun's method by integrating particle positions forward in time from time step $n$ at time $t^n$ to time step $n + 1$ at time $t^{n+1}$  via  

\begin{equation}
x^{n+1}_i = x^{n}_i + \frac{\Delta t^n}{2}\left [ v^n_i + v^{*,n+1}_i \right ]
\end{equation}
Superscripts denote time steps; subscripts, as noted previously, denote particle index.  Here  ${\Delta t^n}$ is the increment of time from the $n$ step to the $n+1$ step, ${\Delta t^n} = t^{n+1} - t^n$.  The intermediate velocity $v^{*,n+1}_i$ is given by 
\begin{equation}
v^{*,n+1}_i=v\left (  x^{*,n+1}_i , t^{n+1}\right ),
\end{equation}
and the intermediate position $x^{*,n+1}$ by
\begin{equation}
x^{*,n+1}_i = x^n_i + \Delta t^n v^n_i .
\end{equation}
The final particle locations $x^{n+1}_i$ and velocities $v^{n+1}_i$ are stored at time $t^{n +1}$, along with hydrodynamic data interpolated to the particle locations, for use in post-processing.

\smallskip

\subsection {Gas-Phase Chemistry}
\label {gasphasechemistry}

The chemical reaction networks  are typically stiff due to the great span of reaction rates involved. Furthermore, astrochemical rates are often highly uncertain; if they are measured at all, they are typically measured at room temperature and then extrapolated down to the lower temperatures typical of GMCs. The physical constraints are combined with the need to flexibly generate networks of hundreds to thousands of reactions required to model the gas-phase chemistry of GMCs. To address these challenges in the gas phase chemistry, we have built a new astrochemistry code framework, ASTROCHEM, upon the chemical kinetics software suite called the Kinetic PreProcessor (KPP) \citep {kpp1, kpp2, kpp3}.

\smallskip
KPP was designed as a general chemical kinetic tool to facilitate the numerical solution of chemical reaction networks. KPP automatically generates FORTRAN code that computes the time-evolution of stiff networks of chemical networks, starting with a high-level specification of the  chemical reaction and rates, and allows the user to select from a variety of stiff integration schemes, including the Livermore Stiff Ordinary Differential Equation Solver (LSODE) package (\citealp {lsode1} and Rosenbrock methods). Significantly, because it symbolically computes the Jacobian and Hessian matrices of the networks needed to interface with numerical integration schemes, it is capable of both generating new network codes, and treating  sensitivities automatically. Furthermore, KPP exploits sparsity in the Jacobian and Hessian matrices to maximize computational efficiency. The ASTROCHEM framework interfaces to existing astrochemical databases and codes, including UMIST \citep {umist06} and Nahoon \citep{wakelamherbst08}. The resulting framework features a high degree of flexibility in creating astrochemical reaction networks and analyzing their  corresponding sensitivities. In the present paper, we focus primarily on carbon, nitrogen, and oxygen ion-neutral and neutral-neutral reactions in the gaseous phase.

\smallskip
The chemical evolution of any species in our chemical network frame can be written in mathematical form as:

\begin{eqnarray}
\frac{\rm d \chi^{\rm i}}{\rm dt}= \sum_{\rm j = 1 \neq i}^{\rm N} {\rm A_r^j a_s^j \chi^j } +  \sum_{\rm k = 1 \neq i}^{\rm N} \sum_{\rm l = 1 \neq i}^{\rm N} {\rm B^{kl} b_s^k c_s^l \chi^k \chi^l } \nonumber \\
- \sum_{\rm m=1\neq i}^{\rm N} {\rm D^{im} d_s^i e_s^m \chi^i \chi^i}  - {\rm E_r^j f_s^j \chi^i } 
\label{generalchemicalevolution}
\end{eqnarray}

\noindent where in the symbol $\chi^{\rm i}$, $\chi$ denotes chemical concentration of a species, superscript i denotes a particular species.
The term $\sum_{\rm j = 1 \neq i}^{\rm N} {\rm A_r^j a_s^j \chi_n^j }$
represents formation of species ${\rm \chi^i}$ from species ${\rm \chi^j}$ through one body dissociation reactions, with ${\rm A_r^j}$ being reaction rate coefficient
for one body dissociation reactions and ${\rm a_s^j}$ is the associated stoichiometric coefficient, and ${\rm N}$ is the total number of species. Similarly, 
${\rm \sum_{\rm k = 1 \neq i}^{\rm N} \sum_{\rm l = 1 \neq i}^{\rm N} {\rm B^{kl} b_s^k c_s^l \chi_n^k \chi_n^l }}$ is the creation term for formation of 
species ${\rm \chi^i}$ from species ${\rm \chi^k}$ and ${\rm \chi^l}$, with ${\rm B^{kl}}$ being two body reaction rate coefficient, ${\rm b_s^k}$ and ${\rm c_s^l}$
are associated stoichiometric coefficients. Reactions of these type may include neutral-neutral, charge transfer, recombination reactions etc. We can account for the destruction reactions too, with ${\rm D^{im}}$ being reaction rate coefficient for two body destruction reactions for species ${\rm \chi^i}$,
${\rm E_r^j}$ is the reaction rate coefficient for single body reactions, ${\rm f_s^j}$ is the stoichiometric coefficient for single body destruction reactions,
${\rm d_s^i}$ and ${\rm e_s^m}$ are the stoichiometric coefficient for two body destruction reactions. 
 
\subsection{Gas-Phase Cooling}
\label {gasphasecooling}

We next describe the gas phase cooling function $\Lambda (n, T)$ adopted in our models. We use nearly-identical cooling terms as those used by \citep {smithrosen03}.  In our model, we include  gas-grain (dust) cooling \citep {hollenbach1989}, where we take the dust temperature to be $ T_{\rm dust} = 10 \ {\rm K} $. We also include cooling due to collisionally-excited vibrational and rotational  modes of molecular hydrogen \citep{leppshull1983}, rotational modes of water \citep{neufelkaufman1993}, and collisionally-excited vibrational modes of water   \citep{hollenbach1989}. We also incorporate cooling due to the dissociation of molecular hydrogen \citep{shapirokang1987,leppshull1983}. Lastly, we include carbon monoxide cooling through several different channels -- including collisionally-excited rotational modes , and collisionally-excited vibrational modes  \citep {neufelkaufman1993,neufeldetal95}. The dominant gas phase molecular cooling mechanism among those listed above is via  CO (carbon monoxide) rotational modes excited by collisions with both atomic and molecular hydrogen, $\Lambda_{\rm COrotcolH} (n, T)$ \citep {neufelkaufman1993,neufeldetal95}. The cooling rate  $\Lambda_{\rm COrotcolH} (n, T)$ for temperatures below 100 Kelvin \citep{neufeldetal95}, while  the cooling rate  $\Lambda_{\rm COrotcolH} (n, T)$ for temperatures above 100 Kelvin is defined by \citep{neufelkaufman1993}. Lastly,  the gas phase cooling function also takes into account the {\it negative} of the hydrogen reformation heating \citep {hollenbachmckee79}. 

\smallskip
When calculating the CO cooling rate, we determine the logarithm of optical depth $\tau_{\rm op}$. 
The optical depth is defined by:
\begin{equation}
{\rm log}_{10}\left ( \tau_{\rm op} \right ) = {\rm log}_{10} \left ( \frac{n_{\rm CO}}{dv/dr} \right ),
\end{equation}
   
\noindent where $n_{\rm CO}$ is the number density of carbon monoxide molecules per cubic centimeter, and $dv/dr$ is the velocity gradient. This is a local approximation which ignores the shielding effect of the cloud and ignores the internal structure of the cloud. We use an average value of velocity gradient calculated over the entire mesh for the optical depth calculation \citep {pan2009} :

\begin{equation}
\frac{dv}{dr} = \frac{1}{3} \left ( \frac{dv_{\rm x}}{dx} + \frac{dv_{\rm y}}{dy} + \frac{dv_{\rm z}}{dz}\right )_{\rm mesh},
\end{equation}

\noindent where the subscript mesh denotes that an averaging over the entire mesh has been done. 

.

\subsection{Shock Modeling}
\label {postprocessing}

\smallskip
When post-processing the Lagrangian tracers, we utilize a one-dimensional model of radiative shock dynamics, including heating, cooling and gas phase chemistry  \citep{hollenbachmckee79,smithrosen03}. The time-dependent equations of hydrodynamics, which respectively represent conservation of mass, momentum, energy and molecular hydrogen are :

\begin{equation}
\frac{\partial \rho }{\partial t} + \frac{\partial (\rho v)}{\partial x} = 0 ,
\label {continuityequation}
\end{equation}  

\begin{equation}
\frac{\partial ( \rho v)}{\partial t} + \frac{\partial (\rho v^2 + p)}{\partial x} = 0 ,
\label {momentumequation}
\end{equation}

\begin{equation}
\frac{\partial ( \rho e )}{\partial t } + \frac{\partial ( \rho e v)}{\partial x } = -p\frac{\partial v}{\partial x} -\Lambda(T,n,f),
\label {energyequation}
\end{equation}

and 
\begin{equation}
\frac{\partial ( f n)}{\partial t} + \frac{\partial (f n v)}{\partial x} = R(T,n,f) - D(T,n,f) .
\label {hydrogenequation}
\end{equation}
Here $n$ is the total number density of hydrogen nuclei, including both atomic and molecular forms of hydrogen, $\rho e$ is the total internal energy density, $f$ is the fractional molecular hydrogen number abundance relative to $n$, $\Lambda$ is  the specific cooling rate, $R$ is the specific rate of formation of molecular hydrogen,  and $D$ is the specific rate of dissociation of molecular hydrogen.

We can express the total number of particle per unit volume $n_{tot}$ in terms of the  fractional helium abundance with respect to hydrogen atomic nuclei $f$(He) and molecular hydrogen abundance as $n_{tot}$ =  $n$[1 + $f$(He) -$f$]. Then, the ratio of specific heats  $\gamma$ and the mean molecular mass $\mu$, taking into account the atomic and molecular phases of hydrogen as well as the atomic helium, may be written as :

\begin {equation}
\quad \gamma = \frac{5.5 -3f}{3.3 - f}, \\
 \quad \mu = \frac{\rho}{n_{tot}m_p}=\frac{1+4 f(He)}{1+f(He)-f}
\end{equation}
Here we have assumed that the gas kinetic temperatures are high enough to excite the rotational degrees of freedom of H$_2$, but not high enough to excite the vibrational degrees of freedom.

This time-dependent hydrodynamic shock model specified by  equations \ref {continuityequation} - \ref {hydrogenequation} are further simplified for solving the hydrodynamic flow in the post-shock region by assuming the shock is stationary. This assumption is valid for the purposes of post-processing because the flow timescale is typically much less than the CFL timestep on the mesh; equivalently, the post-shock cooling layer thickness is much less than the typical grid spacing. In steady state, the Rankine-Hugoniot jump conditions yield the hydrodynamic state immediately post-shock :

\begin{equation}
S = \frac{\rho_{1}}{\rho_{0}} = \frac{v_{0}}{v_{1}}= \frac{(\gamma + 1){\cal{M}}^2}{(\gamma - 1){\cal{M}}^2 + 2}
\end{equation}

\begin{equation}
 \frac{p_1}{p_0} = 1 + ( 1-\frac{1}{S} ){\cal{M}}^2 , \quad \frac{T_1}{T_0} = \frac{p_1}{S p_0} .
\end{equation}
Here $S$ is the compression factor across the shock. Quantities labelled with the subscript ``0''  represent the hydrodynamic state immediately pre-shock, and quantities labelled with the subscript ``1''  are immediately post-shock. Because we consider only non-destructive shocks in this paper, the molecular abundances remain continuous across the shock.

\smallskip
Under the stationarity assumption, the equations of hydrodynamics yield the conservation of mass
\begin{equation} 
\rho v =  \rho_{1} v_{1} ,
\label {conservationmass}
\end{equation}
 and momentum 
 \begin {equation}
p +  \rho v^2  =  p_1 +  \rho_1 v_{1}^2 ,
\label {conservationmomentum}
\end {equation}
in a steady shock relate the density $\rho$, velocity $v$, and pressure $p$ throughout the cooling layer to the immediate post-shock state.  We may rewrite the total internal energy in favor of pressure, $\rho e = p / (\gamma - 1)$, to express the energy equation \ref {energyequation} in the following form :
\begin{equation}
v \frac{\partial p/( \gamma - 1)}{\partial x} + \frac{\gamma}{\gamma - 1}p \frac{\partial v}{\partial x} = -\Lambda(T,n,f).
\label {energyequation2}
\end{equation}

Applying the conservation of mass and momentum (eqns. \ref {conservationmass} and \ref {conservationmomentum}) to equation \ref {energyequation2}, we obtain an ordinary differential equation for the rate of change of the ratio of  specific heats $\gamma$ :
\begin{equation}
\frac{\partial \gamma}{\partial x} = - \frac{4.4}{(3.3 - f)^{2}}\frac{\partial f}{\partial x}.
\label {gammaeqn}
\end{equation}
Similarly, applying conservation of mass and momentum to equation \ref {hydrogenequation}, we obtain another ordinary differential equation governing the composition fraction $f$ of molecular hydrogen  : 
\begin{equation}
\frac{\partial f}{\partial x} = \frac{ {\mu} m_{p}(R-D)}{\rho_1 v_1}.
\label {feqn}
\end{equation}

\smallskip
This formulation yields two simultaneous ordinary differential equations (eqns. \ref {gammaeqn} and \ref {feqn}) governing the post-shock cooling layer, which we solve with the stiff ordinary differential solver LSODE. Our network has over 200 chemical reactions and 66 species (see \ref{chemeqtable}). The electron concentrations, which represent the sole negatively-charged species in our network, are calculated by charge conservation assuming net neutrality. The cooling terms used are same as that used by Smith and Rosen (see \ref{chemeqtable}). We integrate forward through the shock until the post-shock temperature falls to within $0.1 \%$ of the ambient temperature. We then match the post-shock integration condition back onto the Lagrangian tracer trajectory.

\smallskip
At each timestep of the simulation, the Lagrangian tracer particle time history stores the vector position $\vec {x}_i (t)$ and the vector velocity  $\vec {v}_i (t)$ of the $i$th particle in Cartesian coordinates, along with the Eulerian density field $\rho (\vec {x}_i (t))$ interpolated onto the location of the particle. Since we have assumed isothermal gasdynamics for the Eulerian calculation, the Lagrangian tracer particles must  cool back down to the ambient temperature in order to maintain consistency between the Eulerian and Lagrangian fields. 
\smallskip
Due to the highly-intermittent nature of supersonic isothermal turbulence, statistically we expect to find portions of the GMC where the density lies at the extremely rarefied low-end tail of the density probability distribution function (PDF) \citep{vazquez1994,padoanetal1997,passotvazquez1998,federrathetal08}. At these low densities, the local cooling time scale of the gas can be greater than the global dynamical time.In gas parcels of molecular clouds of number density $n$ $\sim{1}$ cm$^{-3}$, the physical cooling time due to rotational line cooling of carbon monoxide collisionally excited by atomic hydrogen is $\sim$ 3 million years,
 assuming $f$(CO) $\sim 7 \times 10^{-5}$. We estimate an upper-bound to the cooling time $t_{\rm cool}$ of the gas  by the ratio of the excess internal energy of the gas above the ambient internal energy $\Delta e$ to the specific cooling rate $\Lambda (n, T)$, taking the collisionally-excited rotational line cooling of carbon monoxide as the dominant cooling mechanism. :

\begin{equation}
t_{\rm cool} \simeq   {\Delta e \over  \Lambda (n, T) } \simeq \frac{\left[1.5n({\rm He}) + 2.5n(H_{2}) \right] \times k_B (T_1 - T_0)}{\Lambda_{\rm COrotcolH} (n, T)} .
\label{coolingtimeestimate}
\end{equation}
When $t_{\rm cool} <\  $shock dynamical time, we identify density jumps as isothermal J-shocks.  When this condition is not satisfied, we integrate forward in time without requiring the temperatures to return to their initial isothermal value.

\begin{figure*}
\centering%
\vspace{-0.5cm}
   \hspace*{2.68cm}\includegraphics[width=140mm]{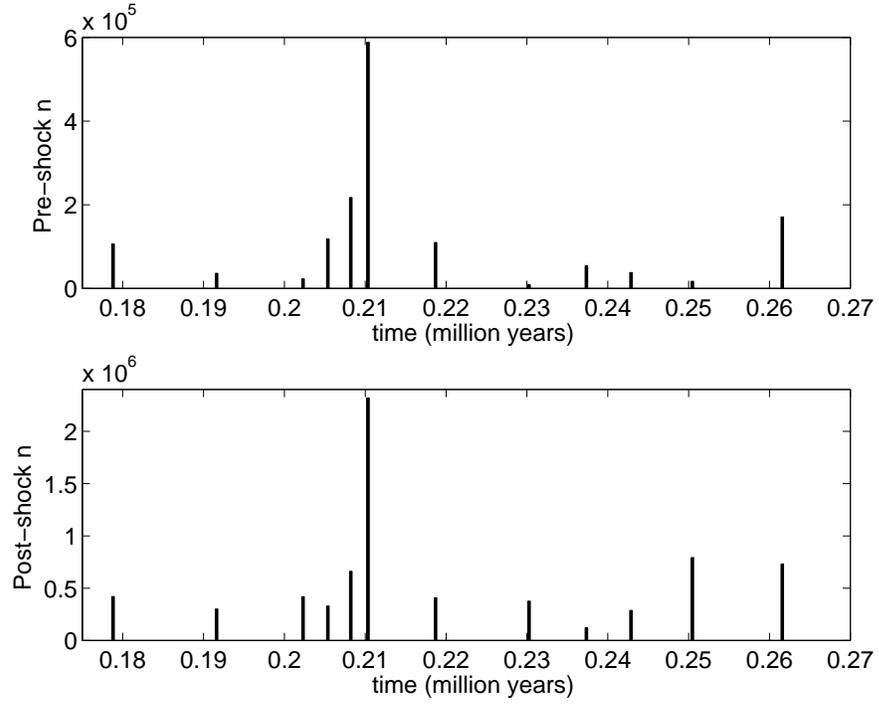}
\caption{Shock location algorithm applied to an individual Lagrangian tracer particle. The top plot shows pre-shock gaseous number density  (in cm$^{-3}$) versus time. The bottom plot shows post-shock gaseous number density  (in cm$^{-3}$) versus time.
\label{traj2}}
\end{figure*}

\smallskip
When postprocessing the trajectory, we filter the density using a boxcar average, to help easily distinguish true shocks, and eliminate small post-shock oscillations. The derivative of filtered density is calculated in a geometrically slope-limited way. A negative time derivative of density represents compression. On finding a continuous region of compression, the maximum physical density inside the region and the density just before entering this region are found. We subject these densities to the  Rankine-Hugoniot shock conditions to retrieve the strength of the shock, assuming this compression to the be effect of a shock. We classify a continuous region with a negative density time derivative and compression ratio greater than 2 as a physical shock. We have plotted figures \ref{traj1} and \ref{traj2} as an example of how the algorithm works. We take a Lagrangian trajectory, and plot the number density variation as a function of time in figure \ref{traj1}. Plotted in another panel in the same figure are the time instants where our algorithm detected a shock, and the vertical axis depicts  the Mach number of the shock our algorithm detected.
In figure \ref{traj2}, the pre-shock and the post-shock densities of the shocks have been plotted  which were shown to be detected  in figure \ref{traj1}. 

\smallskip
We implement a box-car average of the form $ \bar{\rho}^{n} = \lambda \times \bar{\rho}^{n-1} + \left ( 1- \lambda \right )\times {\rho}^{n}$,  where $\bar{\rho}$ is the filtered density and $\rho$ is the unfiltered
 density, and the superscript denotes the time step. Because a sharp discontinuity is diffused on the mesh after filtering, the filtering parameter $ \lambda $ dictates the spatial resolution of shocks. We construct a filter which preserves well-separated distinct shocks and avoids artificially merging them. Our criterion for well-separated distinct shocks is that they must be separated by the inverse Nyquist wavenumber on the mesh, which we take to be 4 grid cells. The relation between the filtering parameter $\lambda$ and the number of cells $N$ over which the discontinuity broadens is given by :
\begin{equation}
\lambda = ( 1 - f_{threshold} )^{\frac{1}{N+1}},
\end{equation}
where $f_{threshold}$ is threshold level, close to unity, which identifies the discontinuity. We set the smoothing parameter $\lambda$ by considering the action of the filter upon a Heaviside function. In all runs described here, we utilize a value of $\lambda=0.3862$, which returns a Heaviside function to a value of $f_{threshold} = 0.99$ over four cells.

\smallskip

\subsection{Solenoidal Heating}
\label {solenoidalheating}
In the previous subsections, we have taken the compressional heating due to shocks into account. We now illustrate the steps taken to account for solenoidal heating due to dissipation in vortex filaments and shear layers. 

\smallskip
The sum total of the spatially-and time-dependent solenoidal and compressional heating per unit mass $\epsilon(x,t)$ on a mesh is represented as \citep{kritsuketal07}:

\begin{equation}
\label {solenoidalheateqn}
\epsilon(x,t)=-\frac{1}{\rm Re} \left(|{\nabla}\times{u}|^2 +\frac{4}{3}|{\nabla}\cdot{u}|^2 \right),
\end{equation}

\noindent where ${\rm Re}$ is the Reynolds number, $|{\nabla}\times{u}|^2$ represents the solenoidal kinetic energy,
$\frac{4}{3} |{\nabla}\cdot{u}|^2$ represents the compressional kinetic energy.

Because our simulations assume the inviscid Eulerian equations of hydrodynamics, we must infer an effective spatially-averaged Reynolds number ${\bar {\rm Re}}$ which treats the numerical and artificial dissipation on the mesh as a real physical viscosity. To do so, we average the dissipation over space,  calculating the ratio of the spatially-averaged sum total of solenoidal and compressional kinetic energy by the  average value of total dissipation $\bar { \epsilon}$,

\begin{equation}
\bar {\rm {Re}} = {\left \langle {|{\nabla}\times{u}|^2+ \frac{4}{3} |{\nabla}\cdot{u}|^2} \right \rangle \over  {\bar\epsilon}}.
\label{meshreynolds}
\end{equation}
\noindent The mean dissipation $\bar {\epsilon}$ is $ \sim { V_{\rm rms}^3 / L_{\rm integral}}$. Using equation \ref {meshreynolds}, we find the averaged Reynolds number of a 512$^3$ simulation of 3-D RMS Mach number 3.5 to be $ \sim 4200 $. 

\smallskip
An important point to consider is the relative strength of the solenoidal and compressional heating. We consider the ratio $ r_{\rm CS}$ of kinetic energy within compressible modes to the total kinetic energy, which is closely related to the ratio of compressible heating to total heating (eqn. \ref {solenoidalheateqn}):  

\begin{equation}
{ r_{\rm CS}}\equiv\frac{\left<|{\nabla}\cdot{u}|^2\right>}{\left<|{\nabla}\cdot{u}|^2\right>+\left<|{\nabla}\times{u}|^2\right>}.
\end{equation}

\noindent The ratio ${ r_{\rm CS}}$ varies between 0.28 \citep {kritsuketal07}  to  0.33 \citep{scaloelm2004a,scaloelm2004b} in simulations in the literature. The  mean value we obtain is  0.32, in close agreement with those previously reported. This ratio is subject to small-scale fluctuations, and hence to resolution effects. Another way to look at this is the ratio between compressive energy and total energy \citep{federrathetal10}, with compressive energy =  $\frac{4}{3}\left<|{\nabla}\cdot{u}|^2\right> $, solenoidal energy = $\left<|{\nabla}\times{u}|^2\right>$, and total energy being equal to sum of solenoidal and compressive energy. This ratio for our case is equal to 0.384. 

\smallskip
Significantly, the ratio $ r_{\rm CS} \simeq 0.32$ indicates that the majority of the turbulent heating within the domain is due to the solenoidal dissipation. This result motivates us to treat the solenoidal dissipation on resolved scales only, using no  subgrid modeling. While we expect the effects of solenoidal heating in our models will not be as pronounced as those works which perform subgrid modelling of vortices \citep{pan2009, godard2009}, the total heating rate is significantly higher than models treating shock heating alone.

\smallskip
We now present the steps for calculating the evolution of the specific internal energy of the gas  change outside of shocks.  We adopt a Lagrangian equation for internal energy evolution outside of shocks, accounting for solenoidal dissipation :

\begin{eqnarray}
\frac{ de}{dt} = { \left [ \epsilon_s (x,t) + {1 \over \rho}\left ( \Gamma_{\rm cr} (n)  - \Lambda  (n, T) \right) \right]}.
\end{eqnarray} 

\noindent Here $de/dt$ is the rate of change of specific internal energy, $\epsilon_s (\vec {x},t)$ is the solenoidal dissipation rate per unit mass at the particle's position ${ \vec {x} (t) }$.   $\Gamma_{\rm cr} (n)$ is the volumetric heating rate due to cosmic ray heating, and $\Lambda (n, T)$ is the volumetric gaseous cooling rate -- all of which are converted from functions of number density $n$ and temperature $T$ to Lagrangian rates per unit mass by dividing by the mass density $\rho$. This internal energy equation is solved separately for each particle using the LSODE stiff implicit solver outside of shocks, along its entire trajectory. 

\smallskip

The volumetric cosmic ray heating in the Eulerian frame due to the kinetic energy released in the ionization of helium and hydrogen molecules is given by:

\begin{equation}
\Gamma^{\rm s}_{\rm cr,V} =   \eta { n} \zeta ( 1 + f {\rm (HE)})  \ {\rm erg} \ {\rm cm}^{-3} \ {\rm s^{-1}},
\end{equation}

\noindent Here, $\eta$ is the kinetic energy released for every ionization electron released, which we take to be 20 ${\rm eV}$  \citep{goldsmith2001}, ${ n}$ is the number density,  $\zeta$ is the ionization rate, $f {\rm (HE)}$ is the relative abundance of helium nuclei compared to hydrogen nuclei.

Using typical values, we scale the cosmic ray heating in the Lagrangian frame :

\begin{eqnarray}
{\Gamma_{\rm cr} \over \rho} = 2.38 \times 10^{-4} \left ( \frac{\eta}{50\ {\rm eV}} \right )\left ( \frac{\zeta}{1.0 \times 10^{-17} \ {\rm s}^{-1}} \right )  \nonumber \\
\left ( \frac{ 1 + f {\rm (HE)}}{1.14} \right ) \left ( \frac{\mu}{2.4} \right )^{-1} \ {\rm erg} \ {\rm g}^{-1} \ {\rm s^{-1}}.
\label{lagrangiancosmicray}
\end{eqnarray}

\noindent The symbols in equation \ref {lagrangiancosmicray} are same as defined earlier. 

\smallskip
Similarly, the Lagrangian heating rate due to  hydrogen molecule reformation $\Gamma_{\rm H_2 \ reform}/ \rho$ is adapted from Hollenbach and McKee \citep {hollenbachmckee79}. 

For comparison, we estimate the solenoidal dissipation from the total turbulent dissipation rate  ${ \epsilon(x,t)}$.  The fiducial scaling of  the total dissipation ${ \bar {\epsilon}}$ used in our work is derived in equation \ref{epsilondissipation}; for a  turbulent RMS Mach 3.5 GMC clump, the mean rate is  $2.02 \times 10^{-4} {\rm erg}  \ {\rm g}^{-1} {\rm s}^{-1}$. Assuming that the solenoidal mode of dissipation is two thirds of the total dissipation, the fiducial value of the average physical solenoidal dissipation ${\bar {\epsilon}_s}$ is then:

\begin{equation}
 { \bar {\epsilon}_s} = 1.34 \times 10^{-4} {\rm erg}  \ {\rm g}^{-1} {\rm s}^{-1}.
\label{solenoidaldissipation}
\end{equation}

On comparing equations \ref {solenoidaldissipation} and  \ref {lagrangiancosmicray}, we find that the dissipation due to the solenoidal on the resolved scales with no subgrid modelling is of the same order of magnitude as the cosmic ray heating on average \citep {pan2009}, though unlike cosmic ray heating, the turbulent dissipation is highly intermittent in space and time. Within the most intense structures of dissipation, the solenoidal  dissipation rate  ${ \epsilon_s(x,t)}$ will exceed the cosmic ray heating rate ${\Gamma_{\rm cr} / \rho} $.

\section {Verification}
\label {verification}

Our shock detection and chemical evolution algorithms depend on the ability of the Lagrangian tracer particles to accurately capture the hydrodynamical evolution. Consequently, the verification of the code modules plays a crucial role in understanding both the successes and the limitations of underlying numerical methods. The FLASH code is the product of nearly a decade of intensive software development, including careful attention to code verification \citep {rosneretal2000}. FLASH is tested nightly against a suite of tests to verify the correctness of core physics modules, including the PPM hydrodynamics module and the Lagrangian tracer module. Standard nightly tests include the Sod shock tube problem \cite {sod78}, the Colella-Woodward colliding blast wave problem \citep {woodwardcolella84}, and the Taylor-Sedov blast wave problem \citep {sedov1946}.  In addition to these standard tests, we also constructed and performed an inclined strong adiabatic shock test to verify the accuracy of Lagrangian tracer particle advection  in the presence of strong shocks inclined with respect to the mesh. 

In addition, the same Eulerian PPM hydrodynamics and Lagrangian tracer modules used here were employed in a large-scale computational study of weakly-compressible turbulence on a 1856$^3$ Eulerian mesh, with  256$^3$ Lagrangian tracer particles \citep {benzietal2008, benzietal2010}. A stringent comparison study of our computational methodology for turbulence was recently completed by five computational and three experimental groups studying fundamental Lagrangian structure functions of turbulent flows \citep {arneodoetal2008}. The excellent agreement obtained in that cross-comparison between both multiple computational codes and experimental results represents both a stringent verification and validation of both the FLASH Eulerian PPM hydrodynamics solver and the Lagrangian tracer module in the subsonic, fully turbulent regime. In a new verification test, detailed below, we consider a rigorous Lagrangian test for fully-developed, isothermal supersonic turbulence by requiring that in a statistically-steady state, the one-point probability distribution functions of the Eulerian and Lagrangian density values must agree with one another. 

Lastly, we also conducted extensive tests verifying our gas phase chemical network solvers and cooling terms.  We compared against previous  results for a static background \citep{iglesias} and single shock chemistry \citep{silk} using ASTROCHEM.

\subsection {Density PDF Verification Test for Supersonic Turbulence }

Moving beyond verification tests in simple geometries, we next consider fully-developed supersonic turbulence. In order to verify our Lagrangian tracer methodology in the regime of fully-developed turbulence, we utilize a key identity; namely, that the one-point Eulerian and Lagrangian PDFs in homogeneous, isotropic turbulence must be identical \citep {pope2000}. Specifically, while both the Eulerian and Lagrangian PDFs derived from the simulation will in general differ from reality, they must both agree with one another by general consistency requirements. In addition, it is important to note that this equivalence between the one-point PDFs holds rigorously, and provides the basis for a rare exact verification test in fully-developed turbulence. 

There are in fact two levels of verification involved in this test.  In the first, exact level, we confirm that the Lagrangian PDF derived from the simulation is statistically identical to the Eulerian PDF, also derived from the same simulation.  Because the Lagrangian tracers follow the mass distribution, this test  is a stringent verification of the Lagrangian evolution. In particular, an arbitrary random distribution of the tracers throughout the spatial domain will not in general succeed in reproducing the Eulerian distribution.  In the second level of the test, we verify the Eulerian PDF against the body of literature of numerical simulations, which have demonstrated that supersonic isothermal turbulence yields an Eulerian mass density probability distribution function which follows a log-normal distribution \citep{vazquez1994, padoanetal1997, vazquez1998, nordlund1999, kritsuketal07, lemasterstone08, federrathetal08, federrathetal10}.  
  
\begin{figure*}
\centering%
\vspace{-0.5cm}
   \hspace*{2.68cm}\includegraphics[width=140mm]{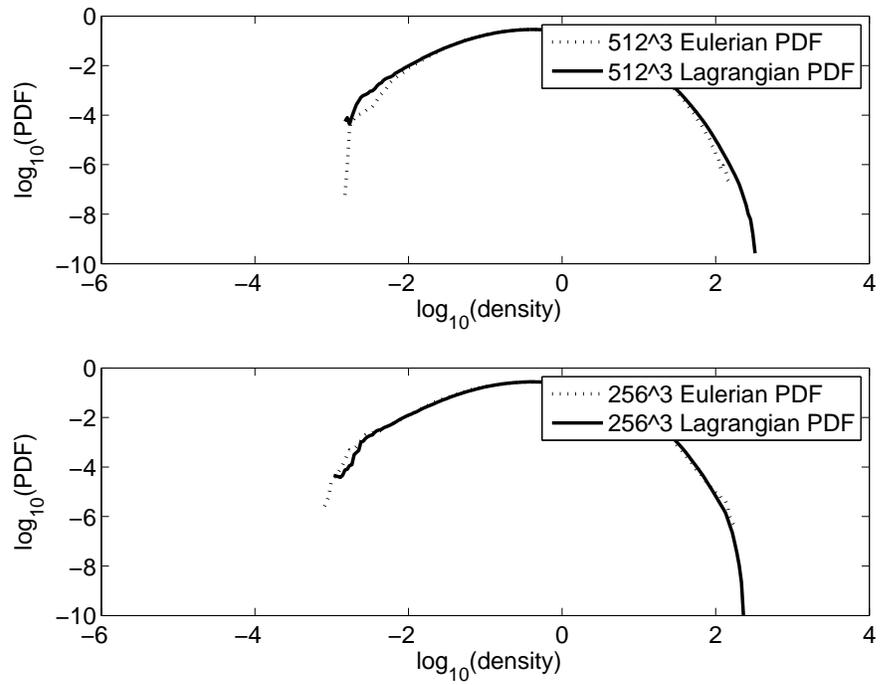}
\caption{Comparison of Lagrangian and Eulerian probability distribution function (PDF) for a 3D-RMS Mach number 3.5, supersonic,
 isothermal, turbulent flow.
\label{verify2}}
\end{figure*}

The Eulerian PDF  is calculated by equally binning the density between the  maximum and minimum values. For the Lagrangian PDF, we weight by mass by weighing the number of particles falling into each density bin by $1 / \rho$, where $ \rho $ is the density of the Lagrangian tracer particle being added to the bin  \citep{kritsuketal07}. We list the important runtime parameters of the simulation in table \ref{verificationtable}. All runs utilize a ratio of specific heats $\gamma = 1 + 10^{-6}$  and an Ornstein-Uhlenbeck  turbulent driving correlation time $ t_{corr} = 0.5 $ in dimensionless units such that the global sound-crossing time is unity. We allow the system to evolve over four dynamical times to reach a steady state.  We confirmed  that the system has in fact reached a steady-state by verifying that the RMS velocity of the system approaches a near-constant value. Time-averages are then performed  over the last two dynamical times. As a demonstration of our calculations where we convince ourselves that Lagrangian and Eulerian PDF come from the same underlying distribution, we have plotted the logarithm of Eulerian and Lagrangian PDF versus logarithm of density for a $256^3$ and $512^3$ simulation in figure \ref {verify2}.

\smallskip
\begin{table}
\caption{Results for Eulerian/Lagrangian PDF Verification.}
\label {verificationtable}
\begin{tabular}{|c|c|c|c|}
\hline
Simulation Parameter &  Run 1 & Run 2 & Run 3 \\
\hline
$N_{grid}^3$ & $ 128^3  $ &   $ 256^3  $    &  $ 512^3  $  \\
\hline
$N_{particle}^3$ & $ 64^3  $ & $ 64^3  $ & $ 64^3  $\\
\hline
$d$-statistic   & 0.0435 &     0.0497 &  0.0435\\
\hline
KS probability &    0.9976  &    0.9866 &  0.9976 \\
\hline
$b$ & 0.4988  & 0.4551 &  0.4351 \\
\hline
\end{tabular}
\end{table}

\smallskip
We perform a best-fit to the time-averaged density probability distribution in log-normal form : 

\begin{equation}
p(\ln \ \rho) \ d(\ln \rho) = \frac{1}{\sqrt(2 \pi \sigma^{2})} \exp \left [ -\frac{1}{2} \left ( \frac{\ln \rho - \overline{\ln \rho}}{\sigma} \right )^{2}\right ] d(\ln \rho)
\label {lognormal}
\end{equation} 
Here the mean of natural logarithm of density is given by
\begin{equation}
\overline{\ln \rho} = - \frac{\sigma^{2}}{2},
\label {lognormal2}
\end{equation}
where ${\rho}$ is the volume density \citep {kritsuketal07,federrathetal08,federrathetal10}. The relation between standard deviation $\sigma$ and Mach number $\cal{M}$ for the parameter $b$ is 
\begin{equation}
\sigma^{2} = \ln\ ( 1 + b^{2}{\cal M}^2) .
\label {lognormal3}
\end{equation}
For our highest resolution simulation, we can obtain the parameters of the fit in two independent ways which are formally identical for an exactly log-normal density distribution. However, the intermittency inherent in turbulence causes the tails of the distribution to differ from an exact log-normal \citep {federrathetal08,federrathetal10}.  In the first parameter determination, we simply determine the mean density directly to find $ \bar{\ln \rho} =  -0.9227 $.  From equation \ref {lognormal2} and equation \ref {lognormal3}, we then obtain $ \sigma = 1.3584 $ and  $b = 0.6596$, in good agreement with previous authors. Alternatively, if we fit the density distribution to equation \ref {lognormal} to obtain $ \sigma = 1.3896 $, we find $b = 0.6938$ from equation \ref {lognormal3}. We would like to point out that value of parameter $ b \sim 1$ for purely compressive forcing and $ b \sim 1/3 $ for  a purely solenoidal forcing \citep {federrathetal08,federrathetal10}. This apparent conflict in the determined values of $b$ points to intermittency inherent in turbulence \citep {federrathetal08,federrathetal10}. 

Both the Eulerian and Lagrangian PDF distributions we obtain  show significant departures from a log-normal fit at low densities. While early low-resolution  studies produced a log-normal fit to the density PDFs centered around the mean,  more recent simulations at higher resolution have tended to be skewed to lower density values \citep{padoanetal1997, kritsuketal07}. The low-density tail is more subject to turbulent intermittency, and so we expect that the actual PDF should skew  to low density values. Kritsuk et al 2007 have conducted perhaps the highest-resolution study of isothermal supersonic turbulence statistics at $ {\cal{ M}} = 6 $  on a $  2048^3 $ mesh, averaged over many snapshots, accumulating the PDF for $ ~ 10^{11} $ data points. The skewness seen in Kritsuk et al's results has since been confirmed by many other grid and SPH codes  \citep {kitsionas}. Moreover, the trend is for higher resolutions to be even less skewed towards low density in the mean, because of the greater sampling of the intermittent tails of the turbulent distributions, which is the case for both grid and SPH based codes  \citep{price}.

\smallskip
We use the Kolmogorov-Smirnov(KS) test to check for the probability that both the Lagrangian and Eulerian probability distribution are drawn from the same  underlying density distribution. The KS test is based upon the
 $d$-statistic, which is defined as the maximum difference in the cumulative distribution functions of the probability distributions being compared. As a consequence, the KS test is insensitive to any binning of the
 underlying PDFs, as well as any presumed functional fit. At our highest resolutions of $512^3$, we find a $d$-statistic of 0.0435, which implies a KS probability that the Lagrangian and Eulerian density PDF originate 
from the same underlying distribution with probability of 0.9976. This result is a strong, rigorous verification of both the Eulerian and the Lagrangian hydrodynamics in a supersonic isothermal turbulent flow.

\section {Results}
\label{resultssection}

\subsection{Time-Evolution of a Single Lagrangian Parcel}
\label {singlelagrangian}

\smallskip
\begin{table}
\begin{center}
\caption{List of chemical species evolved in the model}
\label {tab:speclist}
\begin{tabular}{c c c c c c  }
\hline
{\rm NO$_2$}  & {\rm NH$^{+}$} & {\rm CO$_2^{+}$} & {\rm H$_2^{+}$} & {\rm HCN$^{+}$} & {\rm NH$_2^+$}  \\
{\rm HNCO} & {\rm CH$^+$} & {\rm OH$^+$} & {\rm N$^+$} & {\rm O$^+$} & {\rm CO$^+$}   \\
{\rm He} & {\rm HCO$_2^+$} & {\rm HCO} & {\rm HNO} & {\rm NH$_4^+$} & {\rm H$_3$CO$^+$}   \\
{\rm CN$^+$} & {\rm HN$_2^+$} & {\rm NO$^+$} & {\rm H$_2$CO$^+$} & {\rm H$_2$CN$^+$} & {\rm CH$_2$}   \\
{\rm CH$_4$} & {\rm NH$_3^+$} & {\rm HCN} & {\rm NH} & {\rm H$_2$CO} & {\rm NCO}   \\
{\rm CH$_3^+$} & {\rm CH$_3$} & {\rm H$_2$O} & {\rm H$_3$O$^+$} & {\rm NH$_3$} & {\rm NH$_2$}   \\
{\rm OH} & {\rm NO} & {\rm N} & {\rm HCO$^+$} & {\rm CO} & {\rm H$_2$}   \\
{\rm C} & {\rm H$^+$} & {\rm H$_3^+$} & {\rm He$^+$} & {\rm O$_2$} & {\rm C$^+$}  \\
{\rm HNCO$^+$} & {\rm H$_2$NCO$^+$} & {\rm Na$^+$} & {\rm O$_2^+$} & {\rm N$_2$} & {\rm H}  \\
{\rm e$^-$} & {\rm NCO$^+$} & {\rm N$_2^+$} & {\rm CH$_2^+$}  & {\rm H$_2$O$^+$} & {\rm CO$_2$} \\
& {\rm NH} & {\rm Na} & {\rm O}  & {\rm CN} &\\
\hline
\end{tabular}
\end{center}
\end{table}

\begin{figure*}
\centering%
\vspace{-0.5cm}
   \hspace*{2.68cm}\includegraphics[width=140mm]{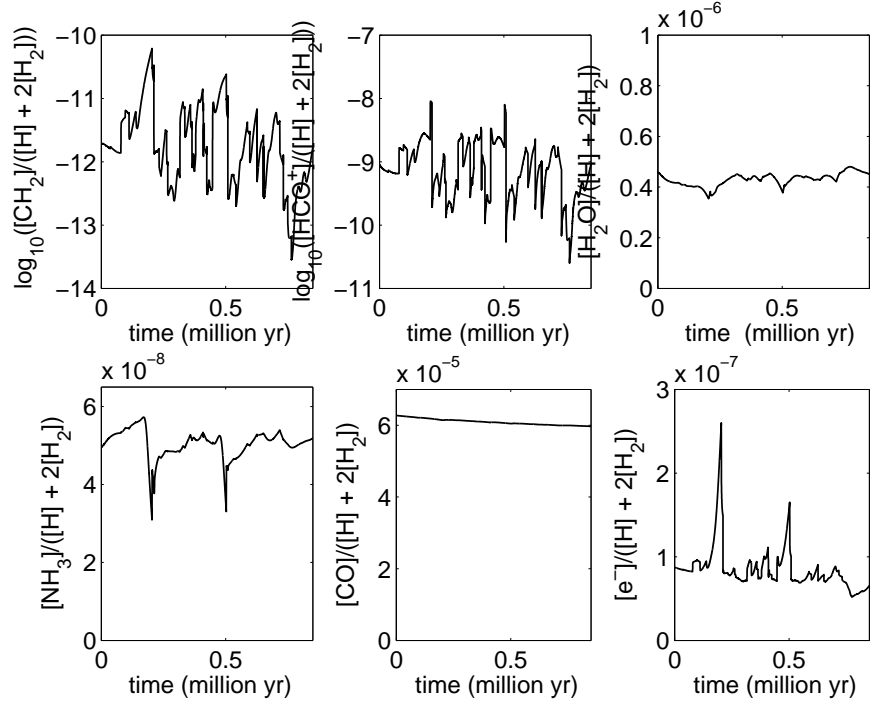}
\caption{ Variation of normalized abundances of various species, relative to the total hydrogen nuclei density $(= \rm{n_{H} + n_{H_2}})$, shown versus time for an individual Lagrangian tracer particle. Part 1
\label{conc1}}
\end{figure*}

We present the detailed chemical evolution along one tracer particle trajectory, under conditions typical of GMC clumps. The initial state is in chemical equilibrium at the background state. We begin the simulation in chemical equilibrium on the uniform background state of the cloud. Changes in chemical concentrations come about due to the density fluctuations in the background cloud, as well as  the temperature enhancements in 
s which a gas parcel encounters. The tracer particle is taken from a $512^3$ resolution Eulerian simulation, with an adiabatic coefficient of $\gamma = 1 + 10^{-6}$, in a supersonic turbulent medium, with a 3D RMS Mach number of 3.5. We  post-process Lagrangian trajectories as described in section  \S \ref {postprocessing} in calculating the chemical evolution for our chemical network. The complete list of chemical species evolved in our simulation are listed in the table \ref{tab:speclist}. The time evolution of the relative concentration of some species along a single Lagrangian tracer are shown in figure \ref {conc1}.  The initial chemical concentrations, relative to the total number of hydrogen nuclei, used in deriving an initial chemical equilibrium at a fixed background number density of $2 \times 10^{4} \ {\rm cm}^{-3}$ and ambient interstellar temperature of 10 Kelvin are listed in table \ref {tab:conc_table}.

\smallskip
Inspection of  figure \ref {conc1}  reveals that some species  (e.g., ${\rm CH_2}$, ${\rm HCO^+}$) experience rapid changes in relative chemical concentrations, in some cases of up to three orders of magnitude compared to their relative concentration in a static  medium. Other species (e.g., ${\rm H_2O}$ and ${\rm NH_3}$) evolve more smoothly, over a dynamic timescale. Still others (e.g., CO) remain relatively constant.

\smallskip

To better understand the chemical evolution of these species, we identify two characteristic time scales, which jointly determine the chemical and thermodynamic evolution in the post-shock cooling layer. The first time scale is the reaction time, which is defined for a given species $X$ to be the inverse of the  the physical rate of change of concentration of the species, normalized to its own number density : $ t_{\rm chem} (X) = {n_{\rm X} / ({d n_{\rm X} / dt}) } $. Here $n_{\rm X}$, or $[{\rm X}]$ is the number  density of a given species $X$. The second time scale is the cooling time defined as : $ t_{\rm cool} = {T}/({dT /dt}) $, the time in which gas phase temperature reaches the background temperature of the molecular cloud.

\smallskip
When computing the cooling time, we have taken the dominant cooling rate to be that due to CO rotational lines excited by collisions with atomic  and molecular hydrogen. \noindent Assuming molecular hydrogen and helium to be the dominant species, the internal energy required for a parcel of gas in the immediate post-shock temperature $T_{\rm p}$ to drop  down to the ambient temperature $ T_{\rm ambient}$ is $(1.5 [{\rm He}] + 2.5 [{\rm H_2}] ) \times k_{\rm B} (T_{{\rm p}} - T_{{\rm ambient}}))$, while the cooling rate is given by $\Lambda (n, T)$ (see section \ref {gasphasecooling}).  The cooling time in the post-shock flow is then given by 

\begin{equation}
t_{\rm cooling} = \frac{(1.5 [{\rm He}] + 2.5[{\rm H}_{2}]) \times k_{\rm B} (T_{\rm p} - T_{\rm ambient}))}{\Lambda (n, T)}.
\end{equation}
This cooling time is the same for all the chemical species.

\smallskip
The chemical time $t_{\rm chem}$ and the cooling time $t_{\rm cooling}$ can be used to define a {\it local} Damk\"{o}hler number ${\rm Da(X)}$ in the post-shock flow for a given species ${\rm X}$ as
\begin{equation}
{\rm Da(X)} = \frac{t_{\rm cooling}}{t_{\rm chem}({\rm X})}.
\label{damchemdef}
\end{equation}

Significantly, the local post-shock Damk\"{o}hler number of a given species specifies whether the chemistry of that species is either fast or slow, relative to the cooling time in the post-shock flow. Specifically, for Da $> >1$, the chemical timescale $t_{{\rm chem}}({\rm X})$ is rapid in comparison to the post-shock cooling time $t_{\rm cooling}$. The chemical evolution of such large Da species is highly sensitive to the presence of shocks.  In contrast, for Da  $< <1$,  the chemical timescale is slow in comparison to the cooling time. In effect, these low Da species do not ``see'' the shocks, and  consequently, the chemical evolution of these species are insensitive to the presence of shocks.

\smallskip
\begin{table}
\begin{center}
\caption{Number Abundances Relative to Hydrogen Nuclei}
\label {tab:conc}
\begin{tabular}{c c}
\hline
{\rm Element}  & $ \quad \quad {\rm Relative \ Number\ Abundance} $ \\
\hline
{\rm Atomic hydrogen H} $\dotfill$ & 0.4 \\
{\rm Molecular hydrogen H$_2$} $\dotfill$ & 0.3 \\
{\rm Helium He} $\dotfill$ & 0.14 \\
{\rm Atomic oxygen O} $\dotfill$ & $ 1.17 \times 10^{-4}$   \\
{\rm Atomic carbon C} $\dotfill$ & $ 1.47 \times 10^{-5}$ \\
{\rm Atomic sodium Na} $\dotfill$ & $1.02 \times 10^{-7}$ \\
{\rm Atomic nitrogen N} $\dotfill$ & $2.14 \times 10^{-5}$ \\
{\rm Carbon monoxide CO} $\dotfill$ & $ 5.87\times 10^{-5} $ \\
\hline
\end{tabular}
\end{center}
\end{table}

\begin{figure*}
\begin{center}
\caption{ The top plot shows the variation in normalized $\rm{H_2O}$, $\rm{CO}$ and $\rm{CH_2}$ abundance versus time for a single Lagrangian tracer, in the presence of multiple shocks. The bottom plot shows the logarithm of local Damk\"{o}hler number ${\bf Da} $ versus time. See text for definitions.
\label{damall}}
\vspace{-0.5cm}
   \hspace*{2.68cm}\includegraphics[width=140mm]{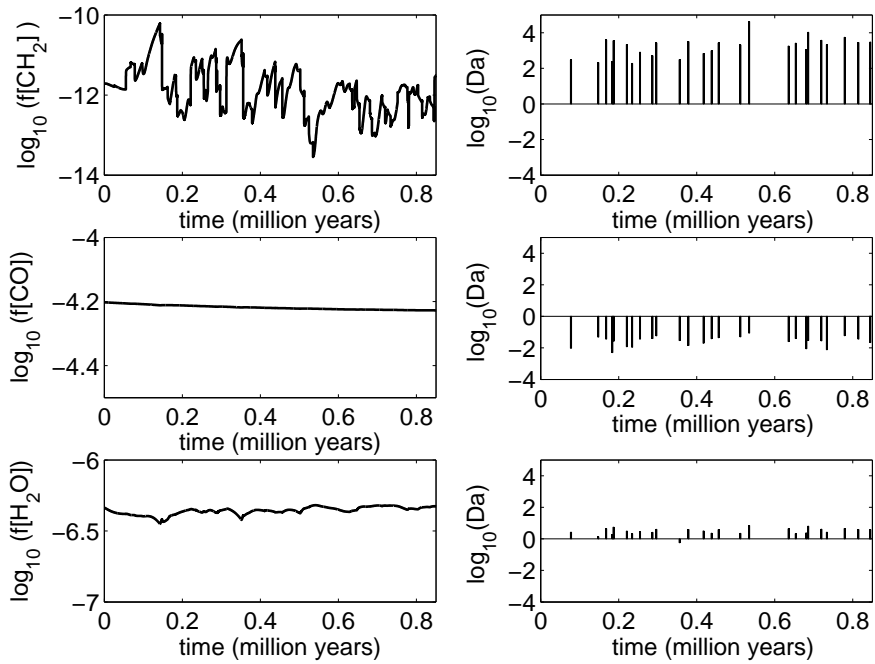}
\end{center}
\end{figure*}

\smallskip
We now directly compare three representative species -- methylene radical (CH$_2$), water (H$_2$O), and carbon monoxide (CO). We calculate the cooling time and chemical time associated with each shock for the sampled trajectory. In figure \ref {damall}, we show results for methylene radical ${\rm CH_2}$. Figure \ref {damall} shows that the Damk\"{o}hler number of methylene radical ${\rm Da{(CH_2)}}$ lies in the range ${\rm 1.24 < log(Da(CH_2)) < 4.01}$. The initial chemical concentration of methylene radical relative to the total number of hydrogen nuclei is $1.97 \times 10^{-12}$. As the trajectory evolves, the maximum relative chemical concentration of ${\rm CH_2}$ is $2.28\times 10^{-11}$, while the minimum relative concentration of methylene radical is $1.64 \times 10^{-14}$. We infer that the relative concentration of ${\rm CH_2}$ increases by a maximum factor of $\sim  11.5$ when compared to the initial static equilibrium state, while the relative concentration of ${\rm CH_2}$ decreases by up to a factor of $\sim 1.2 \times 10^2 \times $ when compared to the initial static equilibrium. The time-averaged Damk\"{o}hler number of methylene radical within the immediate post-shock cooling layers is $\overline{{\rm log_{10}(Da(CH_2))}}=2.76$.

\smallskip
In contrast to the large Damk\"{o}hler number of ${\rm CH_2}$ in figure \ref {damall}, we show
the smaller Damk\"{o}hler number of water ${\rm Da(H_2O)}$ in figure \ref {damall}. We determine that ${\rm Da(H_2O)}$ lies in the range ${\rm -0.636 < log(Da(H_2O)) < 0.556 }$ on a logarithmic scale. We also see that the Damk\"{o}hler number associated with any shock for methylene radical is greater than that of water. This means that for any shock, the equilibration time for water is much larger than that of methylene radical. The initial chemical concentration of water relative to the total number of hydrogen nuclei is $3.94 \times 10^{-7}$. As the calculation evolves, the relative chemical concentration of ${\rm H_2O}$ attains a maximum of $4.80 \times 10^{-7}$. The time-averaged Damk\"{o}hler number of water calculated within the immediate post-shock layers is $\overline{{\rm log_{10}(Da(H_2O))}} = -0.02$.

\smallskip
Lastly, some species like
${\rm CO}$ (figure \ref {conc1}), have nearly-constant relative chemical concentrations as the simulation evolves. As an illustrative example of this class of species, we consider CO here. From figure \ref {damall}, we determine that the Damk\"{o}hler number of carbon monoxide ${\rm Da(CO)}$ on a logarithmic scale lies in the range ${\rm -3.02 < log(Da(CO)) < -1.39}$. The relative concentration of carbon monoxide varies between $5.81 \times 10^{-5}$ and $5.85 \times 10^{-5}$, which translates to a fractional change of $\sim 6.8 \times 10^{-3} $ in the relative concentration of carbon monoxide as compared to its static equilibrium state. The time-averaged Damk\"{o}hler number of carbon monoxide within the immediate
post-shock cooling layers is $\overline{{\rm log_{10}(Da(CO))}} = -1.99$. We note that our results for the behavior of carbon monoxide differ from that of \citep {gloveretal10}, as we only consider gaseous phase reactions, excluding external surface chemistry and FUV radiation. The specific conclusions for CO hold for this class of species. Specifically, for species whose chemical time scale is disproportionately larger than the cooling time, there is virtually no change in the relative concentration of the species throughout its evolution

\smallskip
Based on these results, we classify the evolution of chemical species in a supersonic turbulent flow according to their Damk\"{o}hler number. In much the same way as variations in dimensionless numbers characterize the transition between  physical regimes in other hydrodynamic flow problems -- the most notable of course being the transition to fully-developed turbulence characterized by the Reynolds numbers --  we characterize the transition in the character of post-shock chemical evolution of a species by its Damk\"{o}hler number. We list our empirical findings in table \ref {tab:damtable}, where we identify whether the  concentration of a species is either shock-enhanced, slowly-varying, or frozen-in, depending on its Damk\"{o}hler number. The ranges are drawn from the results in this section, and are intended to be rough quantitative guides to the evolutionary regimes.

\begin{table}
\begin{center}
\caption{Chemical Regimes Divided by Damk\"{o}hler number}
\label{tab:damtable}
\begin{tabular}{|c|c|}
\hline
Damk\"{o}hler number & Physical Significance \\
\hline
$ 0.50 \leq {\rm log_{10} Da} $  &  {\rm Shock-enhanced, stochastic variation} \\
\hline
$ -1.50 \leq {\rm log_{10} Da} \leq 0.50 $ &  {\rm Slowly-varying species} \\
\hline
$  {\rm log_{10} Da} \leq -1.50 $  &  {\rm Frozen-in abundance} \\
\hline
\end{tabular}
\end{center}
\end{table}

\subsection{Time-Evolution of an Ensemble of Lagrangian Parcels}

To gather a fuller picture as to the range of variation present in the physical and chemical structure of the GMC clump analyzed in section \ref {singlelagrangian}, in this section we present the aggregate statistics for an ensemble of 100 post-processed particles. First, in order to analyze the thermodynamic structure of the clump, we construct a joint PDF of density and temperature $f (\rho, T)$, defined such that the integral over all temperatures of the joint PDF returns the PDF of the density field :

\begin {equation}
f (\rho) = \int_0^\infty f (\rho, T) dT
\end {equation}
Conversely, integration over density returns the PDF of the temperature field $f (T)$.

\smallskip

\begin{figure*}
\centering%
\vspace{-0.5cm}
   \hspace*{2.68cm}\includegraphics[width=180mm,angle=0]{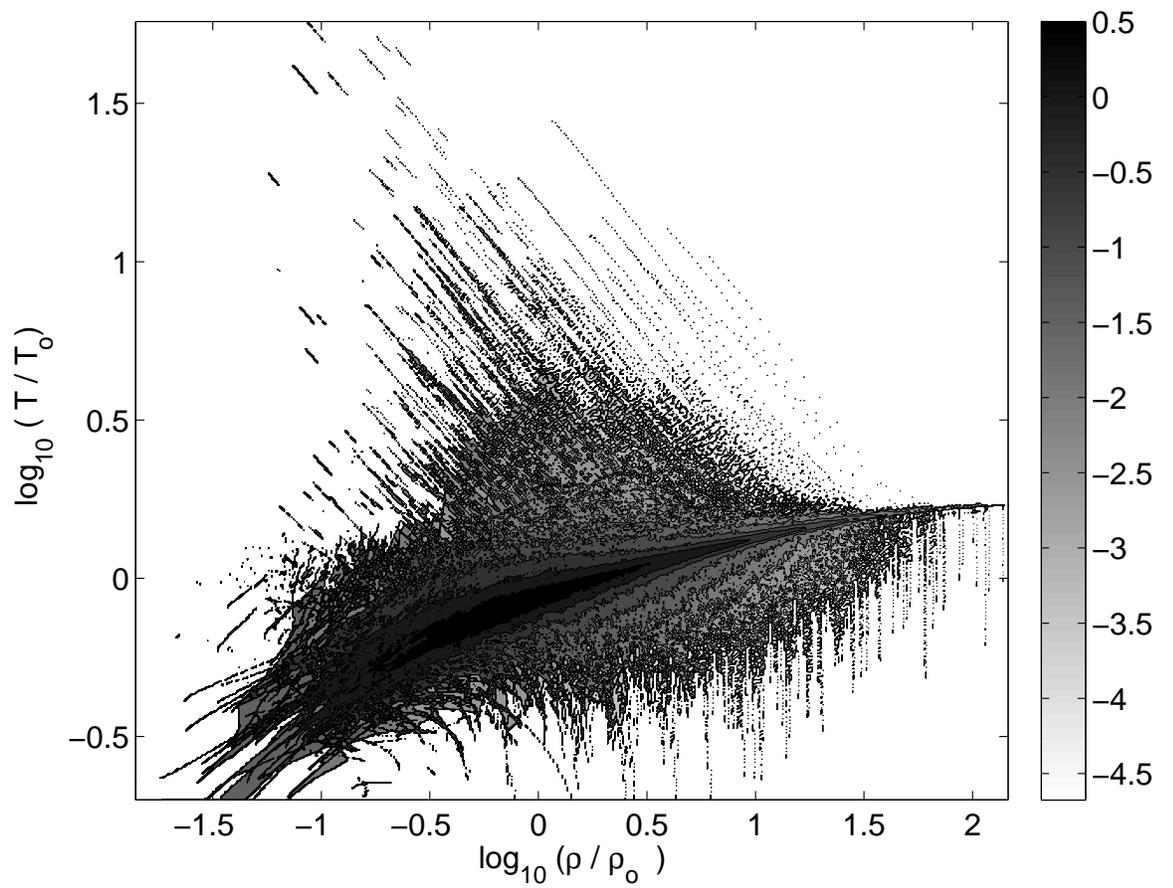}
\caption{The top plot shows contours of joint probability distribution function of density and temperature drawn in a 2-d plane. The temperature is denoted on the vertical axis as $T$, the ambient temperature being 14.8 Kelvin. On the horizontal axis is the density $\rho$, which is normalized by the mean density $\rho_{\rm o}$. Both the quantities are drawn on a logarithmic scale. 
\label{jointpdfcountour1200}}
\end{figure*}

In order to compute the Eulerian joint probability distribution function of density and temperature shown in the figure \ref {jointpdfcountour1200}, we bin the trajectories in a $500 \times 500$ matrix of log density and log temperature, weighing densities by a factor of $1 / \rho$. The entire joint PDF is accumulated by a random sample of 200 particles, averaged over all times.  

\smallskip

Figure  \ref {jointpdfcountour1200} depicts the joint probability distribution function $ f (\rho, T)$. Significantly, because we allow both time-dependent heating and cooling of the gas, the gas temperature is a function of the history of the fluid parcels, and therefore the joint PDF is not distributed along an equation of state prescribed by a single adiabat , as it is in simpler models derived from time-independent thermodynamic equilibrium models of the gas \citep {spaanssilk00}.

\clearpage

\clearpage
\begin{figure*}
\centering%
\vspace{-0.5cm}
   \hspace*{2.68cm}\includegraphics[width=140mm]{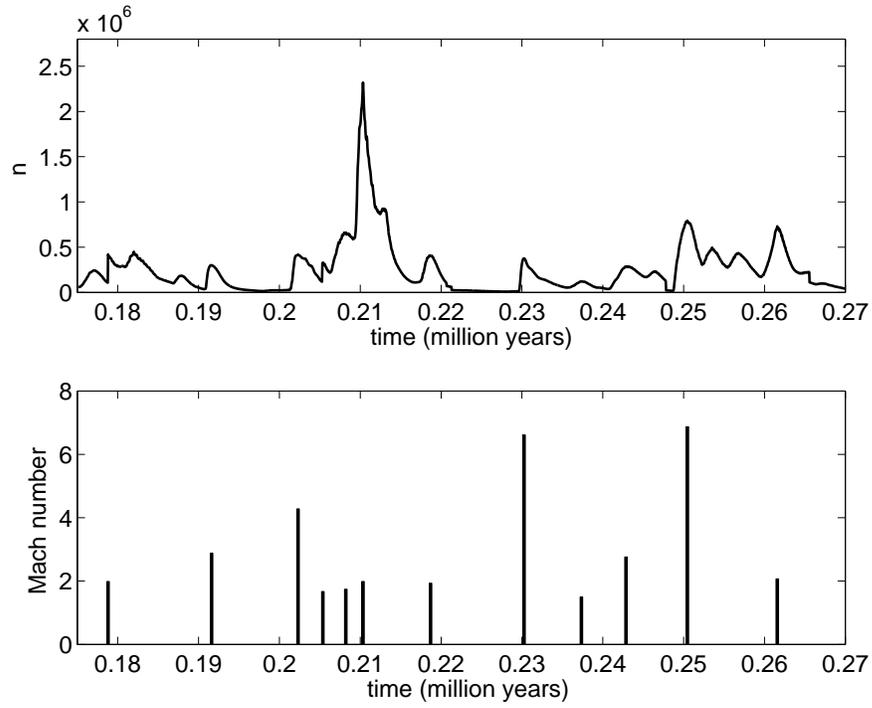}
\caption{ Shock location algorithm applied to an individual Lagrangian tracer particle. The top plot shows gaseous number density (in cm$^{-3}$) versus time. Local Mach number versus time is depicted in the bottom plot.
\label{traj1}} 
\end{figure*}

\clearpage

\clearpage

\clearpage

\clearpage

\clearpage

\smallskip

The joint probability distribution function is a powerful tool which can be used to analyze a number of other questions, including the filling fraction of the warm molecular gas.  By integrating over all densities, we obtain the  temperature probability distribution function $f (T)$.  Using the temperature PDF, we find that $ 2.52 \%$ by volume of the molecular gas  is at temperature of greater than 100 Kelvin. The temperature-integrated density PDF $f (\rho)$ is approximately log-normal. These results are in general accord with the established picture of shock-generated density fluctuations in supersonic turbulence \citep {padoanetal1997}. However,  using our new methodology, we are able to more accurately quantify the temperature structure within thin shock and shear dissipation structures within the clump, a key step forward towards a more realistic chemical model of GMCs.

\smallskip

\section {Discussion and Conclusions}
\label{discussion}

\smallskip
Significantly, a key implication of this work is that the character of the astrochemical evolution of a given species in a fully turbulent medium is robustly determined by its effective  Damk\"{o}hler number, which can be simply estimated, without the need for the complex three-dimensional hydrodynamic simulations, or even detailed shock calculations.  This implication is of significance to observers. For instance, tracers like ${\rm H_2O}$, ${\rm NH_3}$, and ${\rm Na^{+}}$, which are less sensitive to the local turbulent fluctuations within a GMC, and are therefore more reliable tracers of the mean physical state of the cloud itself. Conversely, species like ${\rm CH_2}$, ${\rm HCO^{+}}$, and ${\rm OH}$ are very sensitive to turbulent density fluctuations, and may serve as probes of the dynamics of turbulent intermittency.

\smallskip

These results suggest that an adaptive reduced chemistry model may be feasible in streamlining the computational expense of  a large chemistry network in three-dimensional hydrodynamical and magnetohydrodynamical models. In particular, by estimating both the local cooling time as well as the reaction timescale for a given species, one can determine the Damk\"{o}hler number Da of that species. If Da $>> 1$, then that species can be {\it locally} approximated as being in chemical equilibrium.  Similarly, species for which Da $<< 1$ can be held fixed throughout the post-shock cooling layer.

\smallskip

While in this paper we have treated only hydrodynamic J-shocks,  we plan to extend this model  to include magnetic turbulence, which will enable us to take into account the effects of partial ionization, including ambipolar diffusion. We note that while the normalized electron abundance in these models does not remain constant, it variation is limited to about a factor of 2, from $\sim 0.5 - 1.2 \times 10^{-7}$ (table \ref{tab:conc_table}). This suggests that a relatively straightforward extension of this hydrodynamic methodology to non-ideal magnetohydrodynamics by post-processing the chemical evolution along Lagrangian tracers may be possible.

\smallskip

The non-equilibrium nature of the chemistry of some species may have particular significance for observational and theoretical studies of molecular abundances within star-forming GMC cores, and ultimately protostellar disks and planets. In particular, even in the absence of active star formation and outflows, the background turbulence within the GMC leads to significant departures of the abundances of some species from the chemical equilibrium at the mean cloud density.  This variance in initial concentrations may have a direct influence on the zeroing of chemical clock timescales, which are typically taken about a chemical equilibrium state  \citep {berginetal97, bergintafalla07}. In the context of a fully turbulent origin of GMC cores, a background equilibrium state no longer exists, and the initial core abundances are themselves stochastically established by the shock dynamics. This effect may be particularly pronounced, if in fact GMC cores are formed via shock compression, as some leading models suggest \citep {padoanetal1997}. 

\smallskip

This stochastic variance of chemical abundances in the presence of shocks is in fact hinted at in earlier work by \citep {berginetal98} who studied the evolution of  O$_2$ and H$_2$O under the influence of randomly sequenced C-shocks, as a model of stellar outflows, using Monte Carlo simulations. In their gas-phase models with significantly stronger shock strengths than used here (up to 70 km s$^{-1}$), they found a variation of a factor of $\sim$ 3 in  H$_2$O abundances, and a barely noticeable effect in O$_2$ abundances. Their findings are, however, consistent with ours in that we find both H$_2$O  and  O$_2$ to be low Damk\"{o}hler number species at typical background conditions within J-shocks, and also do not observe large variations in their abundances.  Our results indicate that other species may be more sensitive to shock dynamics, even at the level of the background turbulence in regions away from active stellar outflows.

\smallskip

\smallskip
\thanks{The software used in this work was in part developed by the DOE NNSA-ASC OASCR Flash Center at the University of Chicago. The authors acknowledge fruitful conversations with David Neufeld,
David Falta, and Shivangi Prasad. RF acknowledges research support from NSF Grant CNS-0959382 and AFOSR DURIP Grant FA9550-10-1-0354. This research was supported in part by the National Science Foundation through TeraGrid resources provided by the Louisiana Optical Network Initiative under grant number TG-AST100038.
}

\bibliographystyle{mn2e}
\bibliography{astrochem_paper}

\appendix

\section{Chemical equations}
\label{chemeqtable}

\clearpage
\footnotetext[1]{Reference 1  \citep{silk} \& Reference 2  \citep{umist06} }

\begin{table}
\begin{center}
\caption{Ionization reactions}
\label{tab:hresult1}
\vspace{-0.5em}

{\footnotesize
\begin{tabular}{c c c c} 
\hline\hline 
 \multicolumn{1}{c}{no.} & {Chemical reaction}  &  rate coefficient (s$^{-1}$) & Reference \\ [5ex]
\hline
1) & H$_{2}$ + c.r.p $\rightarrow$ H$_{2}^{+}$ + ${\rm e^{-}}$ + c.r.p$\prime$ $\dotfill$     &  $ 0.95 \times 10^{-17} $  & 1\\
2) & H$_{2}$ + c.r.p $\rightarrow$ H +  H$^{+}$ + ${\rm e^{-}}$ + c.r.p$\prime$ $\dotfill$    & $ 0.05 \times 10^{-17}$ & 1\\
3) & He + c.r.p $\rightarrow$   He$^{+}$ + ${\rm e^{-}}$ + c.r.p$\prime$ $\dotfill$   & $ 1 \times 10^{-17} $  & 1\\
\hline
\end{tabular}}
\end{center}
\end{table}

\begin{table}
\caption{Radiative Association}
\label{tab:hresult11}
\centering 
{\footnotesize
\begin{tabular}{c c c c} 
\hline\hline 
 \multicolumn{1}{c}{no.} & {Chemical reaction}  &  rate coefficient(cm$^{3}$s$^{-1}$) & Reference \\ [5ex]
\hline
4) & C$^{+}$ + H$_{2}$ $\rightarrow$   CH$_{2}$ $^{+}$ + h$\nu$  $\dotfill$    & $ 0.65 \times 10^{-17} $ & 2\\
\hline
\end{tabular}}
\end{table}

\clearpage

\begin{table}
\caption{Positive Ion Molecule reactions reactions part 1 }
\label{tab:hresult2}
\centering 
{\footnotesize
\begin{tabular}{c c c c} 
\hline\hline 
 \multicolumn{1}{c}{no.} & {Chemical reaction}  &  rate coefficient($10^{-9}$cm$^{3}$s$^{-1}$) & Reference\\ [0.5ex]
\hline
5)   &  H$_{2}$ +  H$_{2}^{+}$ $\rightarrow$ H$_{3}^{+}$ + H   $\dotfill$      &  2.08 & 2\\
6)   &  H$_{2}$ +  CO$^{+}$ $\rightarrow$ HCO$^{+}$ + H   $\dotfill$      &  0.75 & 2 \\
7)   &  H$_{2}$ +  N$_{2}$$^{+}$ $\rightarrow$ HN$_{2}$$^{+}$ + H   $\dotfill$     &  2.00 & 2\\
8)   &  H$_{2}$ +  He$^{+}$ $\rightarrow$ H$^{+}$ + H +He   $\dotfill$     &  $   3.7 \times 10^{-5} \times \exp{(-35/T)}$       & 2 \\
9)   &  H$_{2}$ +  O$^{+}$ $\rightarrow$ OH$^{+}$ + H    $\dotfill$      &  1.7 & 2 \\
10)   &  H$_{2}$ +  OH$^{+}$ $\rightarrow$ H$_2$O$^{+}$ + H    $\dotfill$      &  1.01 & 2\\
11)  &  H$_{2}$ +  H$_{2}$O$^{+}$ $\rightarrow$ H$_{3}$O$^{+}$ + H    $\dotfill$     &  0.64 & 2\\
12)  &  H$_{2}$ +  N$^{+}$ $\rightarrow$ NH$^{+}$ + H    $\dotfill$     &  $ \exp{ (-85/T)  }  $ & 2\\
13)  &  H$_{2}$ +  NH$^{+}$ $\rightarrow$ NH$_2$$^{+}$ + H    $\dotfill$    &  $ 1.28  $ & 2\\
14)  &  H$_{2}$ +  NH$_2$$^{+}$ $\rightarrow$ NH$_3$$^{+}$ + H    $\dotfill$      &  $ 0.27  $ & 2\\
15)  &  H$_{2}$ +  CH$^{+}$ $\rightarrow$ CH$_2$$^{+}$ + H    $\dotfill$      &  $ 1.20  $ & 2\\
16)  &  H$_{2}$ +  CH$_2$$^{+}$ $\rightarrow$ CH$_3$$^{+}$ + H    $\dotfill$      &  $ 1.60  $ & 2 \\
17)  &  H$_{2}$ +  CN$^{+}$ $\rightarrow$ HCN$^{+}$ + H    $\dotfill$    &  $ 1.00 $ & 2 \\
18)  &  H$_{2}$ +  HCN$^{+}$ $\rightarrow$ H$_2$CN$^{+}$ + H    $\dotfill$      &  $ 0.9 $ & 2 \\
19)  &  H$_{2}$ +  CNO$^{+}$ $\rightarrow$ HNCO$^{+}$ + H    $\dotfill$      &  $ 1.0  $ & 1\\
20)  &  H$_{2}$ +  HNCO$^{+}$ $\rightarrow$ H$_2$NCO$^{+}$ + H    $\dotfill$     &  $ 1.0  $ & 1 \\
21)  &  He$^{+}$ +  CO $\rightarrow$ C$^{+}$ + O + He    $\dotfill$   &  $ 1.6  $ & 2\\
22)  &  He$^{+}$ +  N$_2$ $\rightarrow$ N$^{+}$ + N + He    $\dotfill$     &  $ 0.96  $ & 2 \\
23)  &  He$^{+}$ +  N$_2$ $\rightarrow$ N$_2$$^{+}$  + He    $\dotfill$      &  $ 0.64  $ & 2\\
24)  &  He$^{+}$ +  O$_2$ $\rightarrow$ O$^{+}$ + O + He    $\dotfill$     &  $ 1.0  $ & 2\\
25)  &  He$^{+}$ +  O$_2$ $\rightarrow$ O$_2$$^{+}$  + He    $\dotfill$     &  $ 3.3 \times 10^{-2}  $ & 2 \\
26)  &  He$^{+}$ +  CN    $\rightarrow$ C$^{+}$ + N + He    $\dotfill$   &  $ 0.88 \times (T/300)^{-0.5} $ & 2\\
27)  &  He$^{+}$ +  NO    $\rightarrow$ N$^{+}$ + O + He    $\dotfill$    &  $ 1.4  $ & 2 \\
28)  &  He$^{+}$ +  NCO   $\rightarrow$ NCO$^{+}$ + He     $\dotfill$     &  $ 1.0  $ & 2 \\
29) &  He$^{+}$ +  NCO   $\rightarrow$ CO$^{+}$ + N + He  $\dotfill$     &  $ 3.0 $ & 2\\
30) &  He$^{+}$ +  NCO   $\rightarrow$ N$^{+}$ + CO + He  $\dotfill$     &  $ 3.0  $ & 2 \\
31)  &  He$^{+}$ +  HNCO  $\rightarrow$ NCO$^{+}$ + H + He $\dotfill$      &  $ 1.0  $ & 1 \\
32)  &  He$^{+}$ +  HNCO  $\rightarrow$ HNCO$^{+}$  + He $\dotfill$     &  $ 1.0  $ & 1 \\
33)  &  He$^{+}$ +  CO2  $\rightarrow$ CO$_2$$^{+}$ + He $\dotfill$      &  $ 0.121  $ & 2  \\
34)  &  He$^{+}$ +  CO2  $\rightarrow$ CO$^{+}$ + O + He $\dotfill$     &  $ 0.87 $  & 2 \\
35)  &  He$^{+}$ +  CO2  $\rightarrow$ O$_2$ + C$^{+}$ + He $\dotfill$      &  $ 0.04  $ & 2 \\
36)  &  He$^{+}$ +  CO2  $\rightarrow$ CO + O$^{+}$ + He $\dotfill$    &  $ 0.1  $ & 2 \\
37)  &  He$^{+}$ +  CO2  $\rightarrow$ O$_2$$^{+}$ + C + He $\dotfill$      &  $ 0.011  $ & 2 \\
38)  &  C$^{+}$ +  O$_{2}$ $\rightarrow$ CO$^{+}$ + O $\dotfill$    &  $ 0.38  $  & 2\\
39)  &  C$^{+}$ +  OH  $\rightarrow$ H$^{+}$ + CO $\dotfill$      &  $ 2.00  $ & 1 \\
40)  &  C$^{+}$ +  NH  $\rightarrow$ H$^{+}$ + CO $\dotfill$      &  $ 2.00  $ & 1 \\
41)  &  C$^{+}$ +  CH  $\rightarrow$ C$_2$$^{+}$ + H  $\dotfill$      &  $ 0.38 \times (T/300)^{-0.5}   $ & 2 \\
42)  &  C$^{+}$ +  H$_2$O  $\rightarrow$ HCO$^{+}$ + H  $\dotfill$      &  $ 0.9 \times (T/300)^{-0.5}  $ & 2 \\
43)  &  C$^{+}$ +  NH$_2$  $\rightarrow$ H$^{+}$ + HCN $\dotfill$   &  $ 2.0  $ & 1 \\
44)  &  C$^{+}$ +  CH$_2$  $\rightarrow$ C$_2$H$^{+}$ + H $\dotfill$      &  $ 2.0  $ & 1 \\
\hline
\end{tabular}}
\end{table}

\clearpage

\begin{table}
\caption{Positive Ion Molecule reactions reactions part 2 }
\label{tab:hresult3}
\centering 
{\footnotesize
\begin{tabular}{c c c c} 
\hline\hline 
 \multicolumn{1}{c}{no.} & {Chemical reaction}  & rate coefficient($10^{-9}$cm$^{3}$s$^{-1}$) \\ [0.5ex]
\hline
45)  &  C$^{+}$ +  HCN  $\rightarrow$ C$_2$N$^{+}$ + H $\dotfill$    &  $ 3.1 \times (T/300)^{-0.5}   $ & 2 \\
46)  &  C$^{+}$ +  CO$_2$  $\rightarrow$ CO$^{+}$ + CO $\dotfill$   &  $ 1.1  $ & 2 \\
47)  &  C$^{+}$ +  NH$_3$  $\rightarrow$ H$_2$CN$^{+}$ + H $\dotfill$      &  $ 1.75 \times (T/300)^{-0.5}  $ & 2 \\
48)  &  C$^{+}$ +  H$_2$CO  $\rightarrow$ HC$_2$O$^{+}$ + H $\dotfill$     &  $ 2.0  $  & 1\\
49)  &  HCO$^{+}$ +  OH  $\rightarrow$ HCO$_2$$^{+}$ + H $\dotfill$     &  $ 1.00 \times (T/300)^{-0.5}   $ & 2  \\
50)  &  HCO$^{+}$ +  H$_2$O  $\rightarrow$ H$_3$O$^{+}$ + CO $\dotfill$     &  $ 2.5 \times (T/300)^{-0.5}   $ & 2 \\
51)  &  HCO$^{+}$ +  NH  $\rightarrow$ NH$_2$$^{+}$ + CO $\dotfill$      &  $ 0.64 \times (T/300)^{-0.5}  $ & 2 \\
52)  &  HCO$^{+}$ +  NH$_2$  $\rightarrow$ NH$_3$$^{+}$ + CO $\dotfill$     &  $ 0.89 \times (T/300)^{-0.5}  $ & 2  \\
53)  &  HCO$^{+}$ +  NH$_3$  $\rightarrow$ NH$_4$$^{+}$ + CO $\dotfill$      &  $ 2.2 \times (T/300)^{-0.5}  $ & 2 \\
54)  &  HCO$^{+}$ +  H$_2$CO  $\rightarrow$ H$_3$CO$^{+}$ + CO $\dotfill$      &  $ 3.3 \times (T/300)^{-0.5} $ & 2  \\
55)  &  H$_3$$^{+}$ +  O  $\rightarrow$ OH$^{+}$ + H$_2$ $\dotfill$      &  $ 0.84  $ & 2 \\
56)  &  H$_3$$^{+}$ +  C  $\rightarrow$ CH$^{+}$ + H$_2$ $\dotfill$     &  $ 2.0  $ & 1 \\
57)  &  H$_3$$^{+}$ +  CO  $\rightarrow$ HCO$^{+}$ + H$_2$ $\dotfill$      &  $ 1.7  $ & 2 \\
58)  &  H$_3$$^{+}$ +  N$_2$  $\rightarrow$ HN$_2$$^{+}$ + H$_2$ $\dotfill$    &  $ 1.8  $ & 2  \\
59)  &  H$_3$$^{+}$ +  OH  $\rightarrow$ H$_2$O$^{+}$ + H$_2$ $\dotfill$      &  $ 1.3 \times (T/300)^{-0.5} $ & 2  \\
60)  &  H$_3$$^{+}$ +  CN  $\rightarrow$ HCN$^{+}$ + H$_2$ $\dotfill$    &  $ 2.0 \times (T/300)^{-0.5}  $ & 2 \\
61)  &  H$_3$$^{+}$ +  H$_2$O  $\rightarrow$ H$_3$O$^{+}$ + H$_2$ $\dotfill$    &  $ 5.90 \times (T/300)^{-0.5}  $ & 2 \\
62)  &  H$_3$$^{+}$ +  CO$_2$  $\rightarrow$ HCO$_2$$^{+}$ + H$_2$ $\dotfill$      &  $ 2.0  $ & 2 \\
63)  &  H$_3$$^{+}$ +  NH$_3$  $\rightarrow$ NH$_4$$^{+}$ + H$_2$ $\dotfill$      &  $ 4.39 \times (T/300)^{-0.5}  $ & 2 \\
64)  &  H$_3$$^{+}$ +  HCN  $\rightarrow$ H$_2$CN$^{+}$ + H$_2$ $\dotfill$     &  $ 8.10 \times (T/300)^{-0.5}  $ & 2 \\
65)  &  H$_3$$^{+}$ +  HCO  $\rightarrow$ H$_2$CO$^{+}$ + H$_2$ $\dotfill$    &  $ 1.7 \times (T/300)^{-0.5} $ & 2 \\
66)  &  H$_3$$^{+}$ +  H$_2$CO  $\rightarrow$ H$_3$CO$^{+}$ + H$_2$ $\dotfill$      &  $ 6.3 \times (T/300)^{-0.5}  $ & 2 \\
67)  &  H$_3$O$^{+}$ +  C  $\rightarrow$ HCO$^{+}$ + H$_2$ $\dotfill$      &  $ 1.0 \times 10^{-2}  $ & 2  \\
68)  &  H$_3$O$^{+}$ +  NH$_3$  $\rightarrow$ NH$_4$$^{+}$ + H$_2$O $\dotfill$    &  $ 2.2 \times (T/300)^{-0.5}  $ & 2  \\
69)  &  H$_3$O$^{+}$ +  HCO  $\rightarrow$ H$_2$CO$^{+}$ + H$_2$O $\dotfill$    &  $ 1.0  $ & 1 \\
70)  &  O$_2$$^{+}$ +  N  $\rightarrow$ NO$^{+}$ + O $\dotfill$      &  $ 0.18  $ & 2 \\
71)  &  HN$_2$$^{+}$ +  CO  $\rightarrow$ HCO$^{+}$ + N$_2$ $\dotfill$      &  $ 0.88  $ & 2 \\
72)  &  CH$_3$$^{+}$ +  O  $\rightarrow$ H$_2$CO$^{+}$ + H $\dotfill$    &  $ 4 \times 10^{-2}  $ & 2  \\
73)  &  CH$_3$$^{+}$ +  N  $\rightarrow$ H$_2$CN$^{+}$ + H $\dotfill$     &  $ 6.7 \times 10^{-2}  $ & 2 \\
74)  &  H$^{+}$ +  CO$_2$  $\rightarrow$ HCO$^{+}$ + O $\dotfill$      &  $ 3.5  $ & 2 \\
\hline
\end{tabular}}
\end{table}

\clearpage

\begin{table}
\caption{ Charge-Transfer Reactions}
\label{tab:hresult4}
\centering 
{\footnotesize
\begin{tabular}{c c c c} 
\hline\hline 
 \multicolumn{1}{c}{no.} & {Chemical reaction}  &  rate coefficient($10^{-9}$cm$^{3}$s$^{-1}$) & Reference\\ [0.5ex]
\hline
75)  &  H$^{+}$ +  O  $\rightarrow$ O$^{+}$ + H $\dotfill$     &  $ 0.731 \times (T/300)^{0.23} \exp{-(225.9/T)} $ & 2 \\
76)  &  H$^{+}$ +  O$_2$  $\rightarrow$ O$_2$$^{+}$ + H $\dotfill$    &  $ 2.00   $ & 2 \\
77)  &  H$^{+}$ +  NO  $\rightarrow$ NO$^{+}$ + H $\dotfill$    &  $ 2.9   $ & 2 \\
78)  &  H$^{+}$ +  OH  $\rightarrow$ OH$^{+}$ + H $\dotfill$   &  $ 2.10 \times (T/300)^{-0.5}  $ & 2  \\
79)  &  H$^{+}$ +  H$_2$O  $\rightarrow$ H$_2$O$^{+}$ + H $\dotfill$      &  $ 6.90  \times (T/300)^{-0.5}  $ & 2 \\
80)  &  H$^{+}$ +  NH$_3$  $\rightarrow$ NH$_3$$^{+}$ + H $\dotfill$   &  $ 3.70  \times (T/300)^{-0.5}  $ & 2 \\
81)  &  H$^{+}$ +  H$_2$CO  $\rightarrow$ H$_2$CO$^{+}$ + H $\dotfill$      &  $ 2.96 \times (T/300)^{-0.5}  $ & 2 \\
82)  &  H$^{+}$ +  CN  $\rightarrow$ CN$^{+}$ + H $\dotfill$    &  $ 1.0  $ & 1 \\
83)  &  H$_2$$^{+}$ +  N  $\rightarrow$ N$^{+}$ + H$_2$ $\dotfill$      &  $ 1.0   $ & 1 \\
84)  &  C$^{+}$ +  NO  $\rightarrow$ NO$^{+}$ + C $\dotfill$   &  $ 0.52  $ & 2 \\
85)  &  C$^{+}$ +  NH$_3$  $\rightarrow$ NH$_3$$^{+}$ + C $\dotfill$    &  $ 2.2 \times (T/300)^{-0.5}  $ & 1 \\
86)  &  C$^{+}$ +  CH$_3$  $\rightarrow$ CH$_3$$^{+}$ + C $\dotfill$    &  $ 1.00   $  & 1 \\
87)  &  O$_2$$^{+}$ +  C  $\rightarrow$ C$^{+}$ + O$_2$  $\dotfill$     &  $ 5.2 \times 10^{-2}   $ & 2 \\
88)  &  O$_2$$^{+}$ +  Na  $\rightarrow$ Na$^{+}$ + O$_2$ $\dotfill$      &  $ 0.71   $ & 2 \\
89)  &  N$_2$$^{+}$ +  Na  $\rightarrow$ Na$^{+}$ + N$_2$ $\dotfill$      &  $ 2.00   $ & 2 \\
90)  &  NO$^{+}$ +  Na  $\rightarrow$ Na$^{+}$ + NO $\dotfill$      &  $ 7.7 \times 10^{-2}   $ & 2 \\
91)  &  H$_2$O$^{+}$ +  Na  $\rightarrow$ Na$^{+}$ + H$_2$O $\dotfill$     &  $ 6.20   $ & 2  \\
92)  &  H$_3$$^{+}$ +  Na  $\rightarrow$ Na$^{+}$ + H$_2$ + H $\dotfill$    &  $ 2.1   $ & 2 \\
93)  &  H$_3$O$^{+}$ +  Na  $\rightarrow$ Na$^{+}$ + H$_2$O + H $\dotfill$      &  $ 3.1   $ & 2 \\
94)  &  HCO$^{+}$ +  Na  $\rightarrow$ Na$^{+}$ + CO + H $\dotfill$     &  $ 2.60   $ & 2 \\
95)  &  HN$_2$$^{+}$ +  Na  $\rightarrow$ Na$^{+}$ + N$_2$ + H $\dotfill$     &  $ 0.50   $  & 1  \\
96)  &  NH$_3$$^{+}$ +  Na  $\rightarrow$ Na$^{+}$ + NH$_3$ $\dotfill$     &  $ 3.2   $ & 2 \\
97)  &  H$_2$CO$^{+}$ +  Na  $\rightarrow$ Na$^{+}$ + H$_2$CO $\dotfill$      &  $ 2.60   $ & 2 \\
98)  &  H$_3$CO$^{+}$ +  Na  $\rightarrow$ Na$^{+}$ + H$_2$CO + H $\dotfill$      &  $ 2.60   $  & 2\\
99)  &  CH$_3$$^{+}$ +  Na  $\rightarrow$ Na$^{+}$ + CH$_3$ $\dotfill$     &  $ 3.40   $  & 2\\
100)  &  H$_2$CN$^{+}$ +  Na  $\rightarrow$ Na$^{+}$ + HCN + H $\dotfill$      &  $ 0.5   $ & 1 \\
101)  &  CN$^{+}$ +  Na  $\rightarrow$ Na$^{+}$ + C + N $\dotfill$     &  $ 0.50   $ & 1 \\
\hline
\end{tabular}}
\end{table}

\clearpage

\begin{table}
\caption{ Ion-Electron Recombination Process}
\label{tab:hresult5}
\centering 
{\footnotesize
\begin{tabular}{c c c c} 
\hline\hline 
 \multicolumn{1}{c}{no.} & {Chemical reaction}  &  rate coefficient($10^{-7}$cm$^{3}$s$^{-1}$) & Reference \\ [0.5ex]
\hline
102)  &  H$^{+}$ +  $e^{-}$  $\rightarrow$ H + $h \nu$   $\dotfill$      &  $ 3.5 \times 10^{-5} (T/300)^{-0.75}   $ & 2 \\
103)  &  Na$^{+}$ +  $e^{-}$  $\rightarrow$ Na + $h \nu$   $\dotfill$      &  $ 2.7 \times 10^{-5} (T/300)^{-0.69} $ & 2  \\
104) &  O$_2$ $^{+}$ +  $e^{-}$ $\rightarrow$ O + O $\dotfill$    &  $ 1.95 \times (T/300)^{-0.70}  $ & 2 \\
105) &  NO$^{+}$ +  $e^{-}$ $\rightarrow$ N + O $\dotfill$     &  $ 4.30 \times (T/300)^{-0.37}   $ & 2 \\
106) &  H$_3$ $^{+}$ +  $e^{-}$ $\rightarrow$ H$_2$ + H $\dotfill$     &  $ 0.234 \times (T/300)^{-0.52}  $ & 2  \\
107) &  H$_3$ $^{+}$ +  $e^{-}$ $\rightarrow$ H + H + H $\dotfill$     &  $ 0.436 \times (T/300)^{-0.52}   $ & 2 \\
108) &  HCO$^{+}$ +  $e^{-}$ $\rightarrow$ CO + H $\dotfill$     &  $ 2.4 \times (T/300)^{-0.69}   $ & 2 \\
109) &  HN$_2$ $^{+}$ +  $e^{-}$ $\rightarrow$ N$_2$ + H $\dotfill$    &  $ 3.6 \times (T/300)^{-0.51}  $ & 2  \\
110) &  H$_3$O$^{+}$ +  $e^{-}$ $\rightarrow$ H$_2$O + H $\dotfill$    &  $ 1.08 \times (T/300)^{-0.50}   $ & 2 \\
111) &  H$_3$O$^{+}$ +  $e^{-}$ $\rightarrow$ OH + H + H $\dotfill$     &  $ 2.58 \times (T/300)^{-0.50}    $ & 2  \\
112) &  NH$_3$$^{+}$ +  $e^{-}$ $\rightarrow$ N$_2$ + H $\dotfill$     &  $ 1.55 \times (T/300)^{-0.50}   $ & 2 \\
113) &  NH$_4$ $^{+}$ +  $e^{-}$ $\rightarrow$ NH$_3$ + H $\dotfill$      &  $ 13.69 \times (T/300)^{-0.5}   $ & 1 \\
114) &  NH$_4$ $^{+}$ +  $e^{-}$ $\rightarrow$ NH$_2$ + 2H $\dotfill$     &  $ 3.19 \times (T/300)^{-0.47}  $ & 2 \\
115) &  H$_2$CN $^{+}$ +  $e^{-}$ $\rightarrow$ HCN + H $\dotfill$      &  $ 10.0 \times (T/300)^{-0.50} $ & 1 \\
116) &  H$_2$CN $^{+}$ +  $e^{-}$ $\rightarrow$ CN + H + H $\dotfill$      &  $ 10.0 \times (T/300)^{-0.50}  $ & 1  \\
117) &  CH$_2$ $^{+}$ +  $e^{-}$ $\rightarrow$ CH + H $\dotfill$     &  $ 1.60 \times (T/300)^{-0.60} $  & 2\\
118) &  CH$_3$ $^{+}$ +  $e^{-}$ $\rightarrow$ CH$_2$ + H $\dotfill$      &  $ 0.775 \times (T/300)^{-0.50}   $  & 2\\
119) &  CH$_3$ $^{+}$ +  $e^{-}$ $\rightarrow$ CH + H + H $\dotfill$      &  $ 2.00 \times (T/300)^{-0.40}   $ & 2 \\
120) &  H$_2$CO $^{+}$ +  $e^{-}$ $\rightarrow$ CO + H + H $\dotfill$     &  $  5.00 \times (T/300)^{-0.50}  $ & 2 \\
121) &  HCO$_2$ $^{+}$ +  $e^{-}$ $\rightarrow$ CO$_2$ + H $\dotfill$    &  $  0.60 \times (T/300)^{-0.64}   $ & 2 \\
122) &  HCO$_2$ $^{+}$ +  $e^{-}$ $\rightarrow$ CO + O + H $\dotfill$      &  $  0.84 \times (T/300)^{-0.64}   $  & 2\\
123) &  H$_3$CO $^{+}$ +  $e^{-}$ $\rightarrow$ H$_2$CO + H $\dotfill$      &  $  2.00 \times (T/300)^{-0.50}   $ & 1 \\
124) &  H$_3$CO $^{+}$ +  $e^{-}$ $\rightarrow$ CO + H + H + H $\dotfill$     &  $  2.00 \times (T/300)^{-0.50}   $ & 1 \\
125) &  CO$_2$ $^{+}$ +  $e^{-}$ $\rightarrow$ CO + O $\dotfill$      &  $  3.80 \times (T/300)^{-0.50}   $ & 2 \\
126) &  CN $^{+}$ +  $e^{-}$ $\rightarrow$ C + N $\dotfill$     &  $  1.80 \times (T/300)^{-0.50}  $ & 2 \\
127) &  H$_2$NCO $^{+}$ +  $e^{-}$ $\rightarrow$ HNCO + H $\dotfill$   &  $ 5.0 \times (T/300)^{-0.50}  $ & 1  \\
128) &  H$_2$NCO $^{+}$ +  $e^{-}$ $\rightarrow$ H$_2$ + NCO $\dotfill$    &  $ 5.0 \times (T/300)^{-0.50}  $ & 1  \\ 
\hline
\end{tabular}}
\end{table}

\clearpage

\begin{table}
\caption{ Neutral-Neutral Reactions}
\label{tab:hresult6}
\centering 
{\footnotesize
\begin{tabular}{c c c c} 
\hline\hline 
 \multicolumn{1}{c}{no.} & {Chemical reaction}  &  rate coefficient($10^{-11}$cm$^{3}$s$^{-1}$) & Reference\\ [0.5ex]
\hline
129) &  O + OH $\rightarrow$ H + O$_2$ $\dotfill$     &  $ 3.5 $ & 2  \\
130) &  O + CH $\rightarrow$ CO + H  $\dotfill$     & $ 6.6  $  & 2  \\
131) &  O + CH$_2$ $\rightarrow$ OH + CH $\dotfill$      & $49.8 \times \exp({-6000/T)} $  & 2  \\
132) &  O + NH $\rightarrow$ NO + H   $\dotfill$     & $11.6$   & 2 \\
133) &  O + NH$_2$ $\rightarrow$ OH + NH  $\dotfill$      & $1.39 \times \exp{(-40/T)} $  & 2  \\
134) &  O + NH$_2$ $\rightarrow$ HNO + H   $\dotfill$     & $4.56 \times \exp({10/T)} $  & 2  \\
135) &  O + CN $\rightarrow$ CO + N $\dotfill$       & $ 4.36 \times (T/300)^{0.46} \exp{(-364/T)} $  & 2   \\
136) &  O + CH$_3$ $\rightarrow$ H$_2$CO + H $\dotfill$      &  $13.0$   & 2 \\
137) &  O + HNO $\rightarrow$ OH + NO $\dotfill$       & $6.0$  & 2  \\
138) &  O + NCO $\rightarrow$ NO + CO $\dotfill$      & $9.43 \times (T/300)^{-0.09} \exp (-100/T)$  & 2  \\
139) &  C + OH $\rightarrow$ CO + H  $\dotfill$       & $10 $  & 2  \\
140) &  C + NO $\rightarrow$ CO + N $\dotfill$      &   $9 $   & 2\\
141) &  C + NH $\rightarrow$ CN + H  $\dotfill$       &  $12 $  & 2 \\
142) &  C + NCO $\rightarrow$ CN + CO  $\dotfill$       &   $10 $ & 2 \\
143) &  N + OH $\rightarrow$ NO + H  $\dotfill$     &  $7.5 \times (T/300)^{-0.18} $ & 2  \\
144) &  N + CH $\rightarrow$ CN + H  $\dotfill$      &  $16.6 \times (T/300)^{-0.09} $ & 2  \\
145) &  N + NH $\rightarrow$ N$_2$ + H $\dotfill$      & $4.98 $  & 2  \\
146) &  N + NO $\rightarrow$ N$_2$ + O $\dotfill$      &  $3.75 \times \exp{(-26/T)} $ & 2  \\
147) &  N + NCO $\rightarrow$ N$_2$ + CO  $\dotfill$       &  $4.0 $  & 1 \\
148) &  H + NCO $\rightarrow$ NH + CO $\dotfill$      &  $12.6 \times \exp{(-515/T)} $ & 2  \\
149) &  CN + O$_2$ $\rightarrow$ NCO + O $\dotfill$     &  $1.86 \times (T/300)^{-0.13}  \times \exp{(40/T)} $ & 2  \\
\hline
\end{tabular}}
\end{table}

\clearpage

\begin{table}
\caption{ Reactions excited by shock heating part 1}
\label{tab:hresult7}
\centering
{\footnotesize
\begin{tabular}{c c c c} 
\hline\hline
 \multicolumn{1}{c}{no.} & {Chemical reaction}  &  rate coefficient(cm$^{3}$s$^{-1}$) & Reference\\ [0.5ex]
\hline
150) &H$_2$ + O $\rightarrow$ OH + H   $\dotfill$   & $ 3.14 \times 10^{-13} (T/300)^{2.7} \exp{(-3150/T)} $ & 2 \\
151) &H$_2$ + OH $\rightarrow$  H$_2$O + H  $\dotfill$   & $ 2.05 \times 10^{-12}(T/300)^{1.52} \exp{(-1736/T)} $ & 1\\
152) &  H$_2$ + H$_2$O $\rightarrow$ OH + H + H$_2$ $\dotfill$   & $5.8 \times 10^{-9} \exp{(-52900/T)} $ & 2   \\
153) &  H$_2$ + C $\rightarrow$ CH + H  $\dotfill$   &  $6.64 \times 10^{-10} \exp{(-11700/T)} $ & 2   \\
154) &  H$_2$ + CH $\rightarrow$ CH$_2$ + H $\dotfill$   & $5.46 \times 10^{-10}  \exp{(-1943/T)} $ & 2 \\
155) &  H$_2$ + CH$_2$ $\rightarrow$ CH$_3$ + H $\dotfill$  &$5.18 \times 10^{-11} (T/300)^{0.17} \exp{(-6400/T)}$ & 2\\
156) &  H$_2$ + CH$_3$ $\rightarrow$ CH$_4$ + H $\dotfill$   & $6.86 \times 10^{-14} (T/300)^{2.74} \exp{(-4740/T)} $ & 2\\
157) &  H$_2$ + N $\rightarrow$ NH + H  $\dotfill$   & $1.69 \times 10^{-9}  \exp{(-18095/T)} $ & 2\\
158) &  H$_2$ + NH $\rightarrow$ NH$_2$ + H  $\dotfill$   & $5.96 \times 10^{-11} \exp{(-7782/T)} $  & 2 \\
159)&  H$_2$ + NH$_2$ $\rightarrow$ NH$_3$ + H $\dotfill$  & $2.05\times 10^{-15} (T/300)^{3.89}\exp{(-1400/T)} $ & 2\\
160) &  H$_2$ + NH$_3$ $\rightarrow$ NH$_2$ + H + H$_2$ $\dotfill$   & $1.5 \times 10^{-8} \exp{(-42400/T)} $ & 1 \\
161) &  H$_2$ + CN $\rightarrow$ HCN + H  $\dotfill$   &  $4.04 \times 10^{-13} (T/300)^{2.87}  \exp{(-820/T)} $ & 2\\
162) &  H$_2$ + NH$_3$$^{+}$ $\rightarrow$ NH$_4$$^{+}$ + H  $\dotfill$   &  $3.36 \times 10^{-14} \exp{(35.7/T)} $ & 2   \\
163) &  H + OH  $\rightarrow$ O + H$_2$   $\dotfill$  &  $6.99 \times 10^{-14} (T/300)^{2.80}  \exp{(-1950/T)} $ & 2  \\
164) &  H + H$_2$O $\rightarrow$ OH + H$_2$   $\dotfill$   & $1.59 \times 10^{-11} (T/300)^{1.20}  \exp{(-9610/T)} $ & 2   \\
165) &  H + CH $\rightarrow$ C + H$_2$   $\dotfill$   &  $1.31 \times 10^{-10} \exp{(-80/T)} $ & 2  \\
166) &  H + CH$_2$ $\rightarrow$ CH + H$_2$   $\dotfill$   & $6.64 \times 10^{-11}  $   & 2  \\
167) &  H + CH$_3$ $\rightarrow$ CH$_2$ + H$_2$  $\dotfill$  & $1.00 \times 10^{-10} \exp{(-7600/T)} $  & 2 \\
168) &  H + CH$_4$ $\rightarrow$ H$_2$ + CH$_3$  $\dotfill$   &  $5.94 \times 10^{-13} (T/300)^{3.00}  \exp{(-4045/T)} $ & 2   \\
169) &  H + NH $\rightarrow$ N + H$_2$   $\dotfill$   &   $1.73 \times 10^{-11} (T/300)^{0.50}  \exp{(-2400/T)} $ & 2 \\
170) &  H + NH$_2$ $\rightarrow$ NH + H$_2$ $\dotfill$   & $5.25 \times 10^{-12} (T/300)^{0.79} \exp{(-2200/T)} $  & 2   \\
171) &  H + NH$_3$ $\rightarrow$ NH$_2$ + H$_2$ $\dotfill$  &  $7.8 \times 10^{-13} (T/300)^{2.40} \exp{(-4990/T)} $  & 2 \\
\hline
\end{tabular}}
\end{table}

\clearpage

\begin{table}
\caption{ Reactions excited by shock heating part 2}
\label{tab:hresult8}
\centering 
{\footnotesize
\begin{tabular}{c c c c} 
\hline\hline 
 \multicolumn{1}{c}{no.} & {Chemical reaction}  &  rate coefficient(cm$^{3}$s$^{-1}$) & Reference\\ [0.5ex]
\hline
172) &  H + H$_2$CO $\rightarrow$ HCO + H$_2$ $\dotfill$   & $4.85 \times 10^{-12} (T/300)^{1.90} \exp{(-1379/T)} $  & 2\\
173) &  H + HNO $\rightarrow$ NO + H$_2$ $\dotfill$   & $4.5 \times 10^{-11} (T/300)^{0.72} \exp{(-329/T)} $ & 2 \\
174) &  H + O$_2$ $\rightarrow$ O + OH $\dotfill$   & $2.61 \times 10^{-10}  \exp{(-8156/T)} $  & 2\\
175) &  O$_2$ + N $\rightarrow$ NO + O $\dotfill$   & $2.26 \times 10^{-12} (T/300)^{0.86} \exp{(-3134/T)} $  & 2 \\
176) &  O$_2$ + NCO $\rightarrow$ CO + NO$_2$ $\dotfill$   & $8.1 \times 10^{-11} \exp{(-773/T)} $ & 1 \\
177) &  O + N$_2$ $\rightarrow$ N + NO $\dotfill$   & $2.51 \times 10^{-10} \exp{(-38602/T)} $ & 2 \\
178) &  O + H$_2$O $\rightarrow$ OH + OH $\dotfill$   &  $1.85 \times 10^{-11} (T/300)^{0.95} \exp{(-8571/T)} $ & 2 \\
179) &  O + HCN $\rightarrow$ OH + CN $\dotfill$   & $6.21 \times 10^{-10} \exp{(-12439/T)} $ & 2   \\
180) &  O + NO $\rightarrow$ N + O$_2$ $\dotfill$   &  $1.18 \times 10^{-11} \exp{(-20413/T)} $ & 2 \\
181) &  O + CH$_4$ $\rightarrow$ OH + CH$_3$ $\dotfill$   & $2.29 \times 10^{-12} (T/300)^{2.20} \exp{(-3820/T)} $ & 2 \\
182) &  O + NH$_3$ $\rightarrow$ OH + NH$_2$ $\dotfill$   &  $1.89 \times 10^{-11} \exp{(-4003/T)} $ & 2 \\
183) &  N + CN $\rightarrow$ N$_2$ + C $\dotfill$   &   $3.00 \times 10^{-10} $ & 2\\
184) &  N + CO$_2$ $\rightarrow$ NO + CO  $\dotfill$   &  $3.20 \times 10^{-13} \exp{(-1710/T)} $  & 2  \\
185) &  OH + CO $\rightarrow$ CO$_2$ + H $\dotfill$  &  $2.81 \times 10^{-13} \exp{(-176/T)} $ & 2  \\
186) &  OH + OH $\rightarrow$ H$_2$O + H $\dotfill$   &  $1.65 \times 10^{-12} (T/300)^{1.14} \exp{(-50/T)} $ & 2 \\
187) &  OH + NH$_3$ $\rightarrow$ NH$_2$ + H$_2$O $\dotfill$  & $1.47 \times 10^{-13} (T/300)^{2.05} \exp{(-7/T)} $  & 2 \\
188) &  OH + NH$_2$ $\rightarrow$ NH$_3$ + O $\dotfill$   & $2.26 \times 10^{-12} (T/300)^{0.76} \exp{(-262/T)} $ & 2 \\
189) &  OH + CH$_4$ $\rightarrow$ H$_2$O + CH$_3$ $\dotfill$   &  $3.77 \times 10^{-13} (T/300)^{2.42} \exp{(-1162/T)}$ & 2  \\
190) &  OH + NO $\rightarrow$ NO$_2$ + H  $\dotfill$   & $5.2 \times 10^{-12} \exp{(-15100/T)} $ & 2    \\
191) &  He$^{+}$ + CH$_4$ $\rightarrow$ CH$_4$$^+$ + He $\dotfill$   & $5.10 \times 10^{-11}  $  & 2  \\
192) &  He$^{+}$ + CH$_4$ $\rightarrow$ CH$_2$$^+$ + He + H$_2$ $\dotfill$   &  $9.5 \times 10^{-10}  $  & 2 \\
193) &  He$^{+}$ + CH$_4$ $\rightarrow$ CH$_3$$^+$ + He + H $\dotfill$   & $8.5 \times 10^{-11}  $ & 2   \\
194) &  He$^{+}$ + CH$_4$ $\rightarrow$ CH$^+$ + He + H$_2$ + H $\dotfill$     &  $2.4 \times 10^{-10}  $  & 2   \\
195) &  He$^{+}$ + CH$_4$ $\rightarrow$ CH$_3$ + He + H$^+$ $\dotfill$   &  $4.8 \times 10^{-10}  $  & 2 \\
196) &  He$^{+}$ + HNO $\rightarrow$ NO + He + H$_+$ $\dotfill$   & $1.0 \times 10^{-9} \times (T/300)^{-0.5} $& 2    \\
197) &  He$^{+}$ + HNO $\rightarrow$ NO$_+$ + He + H $\dotfill$   & $1.0 \times 10^{-9} \times (T/300)^{-0.5} $ & 2   \\
\hline
\end{tabular}}
\end{table}

\begin{table}
\caption{ Reactions catalyzed by dust grains}
\label{tab:hresult9}
\centering 
{\footnotesize
\begin{tabular}{c c c c} 
\hline\hline 
 \multicolumn{1}{c}{no.} & {Chemical reaction}  &  rate coefficient(cm$^{3}$s$^{-1}$) & Reference\\ [0.5ex]
\hline
198) & H + H $\rightarrow$ H$_2$ & 2.5 $\times 10^{-17}$ & 1 \\
\hline
\end{tabular}}
\end{table}

\clearpage
\begin{table}
\begin{center}
\caption{ Average relative concentration data of species}
\label{tab:conc_table}
{\footnotesize
\begin{tabular}{c c c c} 
\hline\hline 
 \multicolumn{1}{c}{species} & {mean relative concentration}  &  standard deviation of relative concentration & initial relative concentration\\ [0.5ex]
\hline
N$^{+}$            &  $   8.530569  \times 10^{-10 } $   &  $     6.529162 \times 10^{-9} $  & $     3.729334  \times 10^{-11} $ \\ \hline
He               &  $   1.398994  \times  10^{-1}$   &  $   4.03005  \times 10^{-5} $  & $     1.4  \times 10^{-1} $ \\ \hline
NH$_4^+$          &  $   1.479286  \times 10^{-11 } $   &  $   1.5110507  \times 10^{-11} $  & $     1.42177  \times 10^{-11} $ \\ \hline
Na$^+$           &  $   6.771889  \times 10^{-8 } $   &  $   5.185934  \times 10^{-9} $  & $     7.872506  \times 10^{-8} $ \\ \hline
NO$^{+}$         &  $   4.384701  \times 10^{-12 } $   &  $   5.8279299  \times 10^{-12} $  & $     6.514852  \times 10^{-12} $ \\ \hline
CH$_2$              &  $   2.671263  \times 10^{-12 } $   &  $   1.518153  \times 10^{-11} $  & $     1.976729  \times 10^{-12} $ \\ \hline
NH$_3^+$          &  $   6.958145  \times 10^{-11 } $   &  $   7.035392  \times 10^{-11} $  & $  6.595174  \times 10^{-11} $ \\ \hline
HCN              &  $   1.282398  \times 10^{-8 } $   &  $   2.189310  \times 10^{-9} $  & $     1.127323  \times 10^{-8} $ \\ \hline
CH               &  $   3.089765  \times 10^{-10 } $   &  $   5.329314  \times 10^{-10} $  & $     2.247113  \times 10^{-10} $ \\ \hline
NCO              &  $   7.930698  \times 10^{-9 } $   &  $   7.993701  \times 10^{-9} $  & $     8.471728  \times 10^{-9} $ \\ \hline
CO$_2$              &  $   9.314529  \times 10^{-10 } $   &  $   7.334628  \times 10^{-10} $  & $     6.407260  \times 10^{-10} $ \\ \hline
N$_2$               &  $   1.017955  \times 10^{-5 } $   &  $   8.040188  \times 10^{-7} $  & $     1.206899  \times 10^{-5} $ \\ \hline
CH$_3^+$          &  $   1.812408  \times 10^{-11 } $   &  $   1.762797  \times 10^{-11} $  & $     2.137332  \times 10^{-11} $ \\ \hline
H$_2$O              &  $   6.833447  \times 10^{-7 } $   &  $   2.910295  \times 10^{-6} $  & $     4.610721  \times 10^{-7} $ \\ \hline
H$_3$O$^+$          &  $   6.044944  \times 10^{-9 } $   &  $   6.726546  \times 10^{-10} $  & $     5.794733  \times 10^{-10} $ \\ \hline
NH$_3$              &  $   5.348606  \times 10^{-8 } $   &  $   1.581378  \times 10^{-8} $  & $     4.958232  \times 10^{-8} $ \\ \hline
NH$_2$              &  $   1.794842  \times 10^{-8 } $   &  $   1.925119  \times 10^{-8} $  & $     1.773805 \times 10^{-9} $ \\ \hline
H                &  $   3.882008  \times 10^{-3 } $   &  $   3.187912  \times 10^{-4} $  & $     5.349509  \times 10^{-3} $ \\ \hline
OH               &  $   2.75309  \times 10^{-8 } $   &  $   5.30112  \times 10^{-8} $  & $     2.047385  \times 10^{-8} $ \\ \hline
NO               &  $   1.519176  \times 10^{-7 } $   &  $   1.552755  \times 10^{-7} $  & $     2.267862  \times 10^{-7} $ \\ \hline
N                &  $   1.99560  \times 10^{-6 } $   &  $   4.254789  \times 10^{-7} $  & $     3.036756  \times 10^{-6} $ \\ \hline
HCO$^{+}$          &  $   8.217715  \times 10^{-10 } $   &  $   7.474217  \times 10^{-10} $  & $     8,544289  \times 10^{-10} $ \\ \hline
CO               &  $   6.038563  \times 10^{-5 } $   &  $   9.499573  \times 10^{-7} $  & $     6.27347  \times 10^{-5} $ \\ \hline
H$_2$               &  $   4.980829  \times 10^{-1 } $   &  $   3.248545  \times 10^{-5} $  & $     4.980191  \times 10^{-1} $ \\ \hline
Na               &  $   3.42078  \times 10^{-8 } $   &  $   5.17178  \times 10^{-9} $  & $     2.327494  \times 10^{-8} $ \\ \hline
CN               &  $   7.377714  \times 10^{-11 } $   &  $   6.538087  \times 10^{-11} $  & $     1.708823  \times 10^{-11} $ \\ \hline
C                &  $   9.874322  \times 10^{-8 } $   &  $   2.569366  \times 10^{-8} $  & $     1.481840  \times 10^{-7} $ \\ \hline
H$^{+}$            &  $   2.651750  \times 10^{-10 } $   &  $   4.785956  \times 10^{-10} $  & $     2.636774  \times 10^{-10} $ \\ \hline
H$_3^{+}$          &  $   1.160548  \times 10^{-9 } $   &  $   1.658192  \times 10^{-9} $  & $     9.953887  \times 10^{-10} $ \\ \hline
He$^{+}$           &  $   5.963769  \times 10^{-10 } $   &  $   1.021091  \times 10^{-9} $  & $     4.988382  \times 10^{-10} $ \\ \hline
O$_2$               &  $   2.297827  \times 10^{-5 } $   &  $   3.048963  \times 10^{-6} $  & $     1.793247  \times 10^{-5} $ \\ \hline
C$^{+}$            &  $   4.391799  \times 10^{-9 } $   &  $   6.69398  \times 10^{-9} $  & $     4.871363  \times 10^{-9} $ \\ \hline
O                &  $   6.863861  \times 10^{-5 } $   &  $   7.060412  \times 10^{-6} $  & $     7.667686  \times 10^{-5} $ \\ \hline
e$^{-}$          &  $    7.665264 \times 10^{-8 } $   &  $   1.741773  \times 10^{-8} $  & $     8.70602  \times 10^{-8} $ \\ \hline
\hline
\end{tabular}}
\end{center}
\end{table}

\clearpage

\section {Microscopic Physics Estimates}
\label {microscopicphysics}

By using the Lagrangian particles, we can cleanly separate the
macromixing due to turbulence from the micromixing due to the
molecular diffusivity, an enormous advantage to this
scheme, in contrast to the Eulerian schemes, in which these two
effects are necessarily intermingled. The section 9.8 of review paper by
\cite{veynante2002} crystallizes the distinction we have made in micromixing and macromixing.
In the present Lagrangian formulation we have neglected the micromixing at the lower scales in
the flow may be modelled as subsonic and incompressible.
We also derive simple estimates for the microphysical transport coefficients.

\smallskip
Next, we estimate the magnitudes of the mass diffusivity coefficient $D$, and the specific turbulent
kinetic energy dissipation rate $\epsilon$, in order to determine the critical turbulent mixing time $(D / \epsilon)^{1/2}$. The neutral-neutral diffusivity ${D_{\rm nn}}$ is given by the product of sound speed and the mean-free path length  of neutral-neutral collisions.
We denote $r_{\rm H_{2}}$ to be the intermolecular
distance between two hydrogen molecules. We denote $\eta^{-}_{\rm H_2}$ is the finite distance at intermolecular
potential is zero. From Lennard-Jones theory, we take $\eta^{+}_{\rm H_2} = 2.5 \eta^{-}_{\rm H_2}$ to be the cut-off
distance beyond which intermolecular forces reduce to zero \citep{haile1997}.
Thus two neutral hydrogen molecules are attracted for $ \eta^{-}_{\rm H_2} < r_{\rm H_{2}} < \eta^{+}_{\rm H_2} $.
The collisional cross-sectional area for neutral-neutral collisions becomes
${\sigma_{\rm nn}} = \pi \left ( \eta^{+}_{\rm H_2} - \eta^{-}_{\rm H_2} \right )^2 $. 
Taking $\eta^{-}_{\rm H_2} = 2.93 \AA$ \citep{curtissbird1964},
$\sigma_{\rm nn} =  {6.07 \times 10^{-15} {\rm cm}^2}  $.

The fiducial value of neutral-neutral diffusivity is given by
\begin{equation}
 {D_{\rm nn} \sim \frac{c_{\rm iso}}{n_{\rm n} \sigma_{\rm nn}}}  \nonumber, 
\end{equation}
\begin{eqnarray}
 \sim 1.00 \times 10^{15} \left( {c_{\rm iso}} \over 0.244 \times 10^4 {\rm \ km \ s^{-1}}\right )  \left ( n_{\rm n} \over {4 \times 10^4 \ {\rm cm}^{-3}} \right ) ^{-1}  \nonumber \\
 \times \left ( \sigma_{\rm nn} \over {6.07 \times 10^{-15} {\rm cm}^2} \right )^{-1} {\rm cm}^2 \ {\rm s}^{-1}
\end{eqnarray} 
 
\noindent where $n_{\rm n}$ is the neutral number density, and ${\sigma_{\rm nn}}$ is the cross-section 
for neutral-neutral $\rm{H_{2}}$ molecular collisions.  

\smallskip
The neutral-neutral dynamic viscosity ${ \mu_{\rm nn} } $ is
\begin{eqnarray}
& \mu_{\rm nn} =  \rho_{\rm n} c_{\rm n} \lambda_{\rm nn} / 2 \nonumber  \\
& \mu_{\rm nn}      =  \rho_{\rm n}  c_{\rm n} /\left ( 2{n_{\rm n} \sigma_{\rm nn}} \right ) ,
\end{eqnarray} 

\noindent where  ${\rho_{\rm n}}$ is the neutral mass density,  $ { \lambda_{\rm nn} }  $ is the mean-free path length for neutral-neutral collisions, and $c_{\rm n}$ is the sound speed for neutral species. 
The factor of $1/2$ appears from angle-averaging over all distances between 0  and $\lambda_{\rm nn}$, and that their average particle velocities change linearly with distance, when  $\lambda_{\rm nn}$ is small. For a completely neutral medium, the kinematic viscosity is given by
\begin{eqnarray}
 \nu_{\rm nn} = \frac{\mu_{\rm nn} }{ \rho_{n} }  = 5.00 \times 10^{14}  \left ( {c_{\rm n}} \over{  0.244 \ {\rm km} \ {\rm s^{-1}}} \right )  \times \left ( {n_{\rm n}} \over {4 \times 10^4  \ {\rm cm}^{-3}} \right )^{-1} \nonumber \\
 \left ( {\sigma_{\rm nn}} \over {6.07 \times 10^{-15}  \ {\rm cm}^{2}} \right )^{-1}  {\rm cm^2} \ {\rm s^{-1}}. 
\label {kinematicviscosityscaling}
\end{eqnarray}
\noindent These estimates have been used in calculating the Reynolds number in section \ref{dimensionlesssubsection}.

\smallskip
The  mean specific turbulent kinetic energy dissipation rate is
\begin{equation}
{\epsilon } 
 =  2.02 \times 10^{-4} \left ( \frac{{\cal M}}{3.5} \right )^3 \left ( {c_{\rm iso}} \over { 0.244  \ {\rm km} \ {\rm s}^{-1}} \right )^{3}  \times \left ( {L} \over {1 \ {\rm pc} } \right )^{-1} {\rm erg}  \ {\rm g}^{-1} {\rm s}^{-1}.
\label{epsilondissipation}
\end{equation}
\noindent While all our estimates here will assume $\epsilon$ is independent of length scale, 
turbulent intermittency results in fluctuations of this value \citep {pope2000}.

\clearpage

\section{Analysis of Damk\"{o}hler Numbers of Representative species}
\label {resultsappendix}

In this appendix, we identify the dominant chemical reactions for the Damk\"{o}hler 
number calculations given in section \S \ref {resultssection}. 

\smallskip
The chemical network used in our codes have been tabulated in Appendix \ref {chemeqtable}. The network consists of both one-body reactions, such as ionization by cosmic ray protons, and two-body reactions, including radiative recombination, charge transfer, recombination and neutral-neutral reactions. We have used the Damk\"{o}hler number to classify the chemical evolution in the post-shock evolution of various species, including ${\rm H_2O}$, ${\rm CH_2}$, and ${\rm CO}$ in section \S \ref {resultssection}. We now derive expressions for these Damk\"{o}hler numbers which we employed in the text.

\smallskip
To begin with, we consider ${\rm H_2O}$. The dominant creation reaction for ${\rm H_2O}$ is
\begin{equation}
110> \ {\rm H_3O^{+} + e^{-}   \rightarrow 0.295 \ H_2O + 0.705 \ OH + 1.705 \ H} \ ; \ k_{110}  \\
\end{equation}

\noindent The numbering of reactions is the same as that in Appendix \ref {chemeqtable}, and k denotes the rate constant of a reaction, with the subscript denoting the reaction number in Appendix \ref {chemeqtable}. This convention is carried forward throughout the Appendix \ref {resultsappendix}. Similarly, the dominant destruction reactions for ${\rm H_2O}$ are

\begin{eqnarray}
42> \ {\rm C^{+} + H_2O     \rightarrow  HCO^{+} + H } \ ; \ k_{42}  \nonumber \\
50> \ {\rm HCO^{+} + H_2O   \rightarrow H_3O^{+} + CO } \ ; \ k_{50} \nonumber\\
61> \ {\rm H_3^{+} + H_2O   \rightarrow H_3O^{+} + H_2} \ ; \ k_{61}  
\end{eqnarray}

\noindent This result is used in figure \ref{damall} and in section \S \ref {resultssection}.

\noindent The rate of change of water at any given instant, taking into account the dominant reactions, is

\begin{eqnarray}
\frac{d[{\rm H_2O}]}{dt} = 0.295 k_{110}[{\rm H_3O^+}][{\rm e^{-}}] - k_{42}[{\rm C^+}][{\rm H_2O}] \nonumber \\
 - k_{50}[{\rm HCO^+}][{\rm H_2O}] - k_{61}[{\rm H_3^+}][{\rm H_2O}]
\end{eqnarray}

\noindent Using the above information, the chemical time scale for water to get to equilibrium is given by
\begin{equation}
t_{\rm chem}\left ({\rm  H_2O }\right ) = \frac{\left [ {\rm H_2O} \right ]}{d\left [ {\rm H_2O} \right ]/dt}
\label{tchemwater}
\end{equation}

\noindent The Damk\"{o}hler number for water ${\rm Da({\rm H_2O})}$ is hence given by (using the definition in
equations \ref {damchemdef}, \ref {tchemwater} and section \ref {gasphasecooling})
\begin{equation}
{\rm Da(H_2O)} = \frac{t_{\rm cooling}}{{\rm t_{chemical}}}
\end{equation}

\smallskip
Next, we consider methylene radical, which we presented as an example of rapidly varying (shock-enhanced)
species in \S \ref {resultssection}. We now derive the expression used for calculating its  Damk\"{o}hler number. The dominant creation reaction for methylene radical ${\rm CH_2}$ is 

\begin{eqnarray}
118>  {\rm CH_{3}^{+} + e^{-} \rightarrow  CH2 + \  H } \ ; \ k_{118} \nonumber \\
\end{eqnarray}

\noindent while the dominant destruction reaction for methylene radical ${\rm CH_2}$ is 

\begin{equation}
166> {\rm H + CH2 \rightarrow CH + H2}  \ ; \ k_{166} 
\end{equation}

\noindent The rate of change of concentration of methylene radical is then given by 

\begin{equation}
\frac{d[{\rm CH2}]}{dt} =  k_{118}[{\rm CH_3^+}][{\rm e^{-}}] - k_{166}[{\rm H}][{\rm CH_2}] 
\end{equation}

\noindent The Damk\"{o}hler number for ${\rm CH_2}$ is then given by 

\begin{eqnarray}
{\rm Da(CH_2)} = \frac{t_{\rm cooling}}{{\rm t_{chemical}}} = \nonumber \\
 \frac{((1.5[\rm{He}] + 2.5[\rm{H_{2}}])\times k_{\rm B} (T_{\rm p} - T_{\rm ambient}))}{\Lambda (n, T)} \nonumber \\
\times \frac{ k_{118}[{\rm CH_3^+}][{\rm e^{-}}] - k_{166}[{\rm H}][{\rm CH_2}]}{\left [ {\rm CH_2} \right ]}
\label{damch2eq}
\end{eqnarray}
 
\noindent We can insert typical values of physical variables in the equation \ref{damch2eq} to get an expression of the  scaling of the Damk\"{o}hler number of ${\rm CH_2}$ under a representative shock.
We select  a shock (a part of the trajectory is shown in figure \ref{traj1}) with Mach number of 3.5.
The post-shock temperature $T_{\rm p}$ is then calculated to be 27.9 K. The actual concentration of each species is indicated in the denominator in the scaled equations which follow below.

\smallskip
To connect the Damk\"{o}hler number to the post-shock temperature, we take the ambient temperature to be $10 \ {\rm K}$. Then the post-shock temperature $T_{\rm p}$ is given by the Rankine-Hugoniot jump condition 
\begin{equation}
\frac{T_{\rm p}}{T_{\rm ambient}} = \frac{\left \{ 2 \gamma {\cal M}^2 - \left ( \gamma - 1 \right ) \right \}  \left \{  \left ( \gamma - 1 \right ){\cal M}^2 + 2 \right \}}{\left ( \gamma + 1 \right )^2 {\cal M}^2}
\label{rankinetempjump}
\end{equation}

\noindent Using the above relation, we can predict the range of the Damk\"{o}hler number for methylene radical as a function of time, provided we know the pre-shock concentration. Significantly, this also implies that the Damk\"{o}hler number is not an universal property, and can vary if  the chemical concentration changes. In the limit of a strong shock we obtain
\begin{equation}
\frac{T_{\rm p}}{T_{\rm ambient}} = \frac{2 \gamma \left ( \gamma -1  \right ){\cal M}^2}{\left ( \gamma + 1 \right )^2}
\end{equation}

\noindent On assuming a completely molecular gas with $\gamma = 7/5$, we obtain
\begin{equation}
\frac{T_{\rm p}}{T_{\rm ambient}} = \frac{T_{\rm p}}{T_{\rm ambient}} = \frac{7 {\cal M}^2}{36}.
\end{equation}

\smallskip
We have presented carbon monoxide as an example of chemically frozen species in \S \ref {resultssection}.
The dominant creation reaction for carbon monoxide ${\rm CO}$ is

\begin{equation}
108> {\rm HCO^{+} + e^{-} \rightarrow CO + H} \ ; \ k_{108} 
\end{equation}

\noindent while the dominant destruction reactions for carbon monoxide ${\rm CO}$ are

\begin{eqnarray}
21> {\rm He^{+} + CO \rightarrow C^{+} + O + He} \ ; \ k_{21} \nonumber \\
57> {\rm H_{3}^{+} + CO \rightarrow HCO^{+} + H_2} \ ; \ k_{57}  
\end{eqnarray}

\noindent Based on the above reactions, the rate of change of carbon monoxide is given by 

\begin{equation}
\frac{d[{\rm CO}]}{dt} = k_{108}[{\rm HCO^+}][{\rm e^{-}}] - k_{21}[{\rm CO}][{\rm H_3^+}] -  k_{57}[{\rm CO}][{\rm He^+}] .
\end{equation}

\noindent Using the above relations, the Damk\"{o}hler number for carbon monoxide is given by 
\begin{eqnarray}
{\rm Da(CO)} = \frac{t_{\rm cooling}}{{\rm t_{chemical}}} = \nonumber \\
 \frac{((1.5[\rm{He}] + 2.5[\rm{H_{2}}])\times k_{\rm B} (T_{\rm p} - T_{\rm ambient}))}{\Lambda (n, T)} \nonumber \\
\times \frac{k_{108}[{\rm HCO^+}][{\rm e^{-}}] - k_{21}[{\rm CO}][{\rm H_3^+}] -  k_{57}[{\rm CO}][{\rm He^+}]}{\left [ {\rm CO} \right ]}
\end{eqnarray}

\label{lastpage}

\end{document}